\documentclass[fleqn,usenatbib]{mnras}

\usepackage{newtxtext,newtxmath}
\usepackage{xfrac} % better fit fractions inline using sfrac
\usepackage[T1]{fontenc}

\DeclareRobustCommand{\VAN}[3]{#2}
\let\VANthebibliography\thebibliography
\def\thebibliography{\DeclareRobustCommand{\VAN}[3]{##3}\VANthebibliography}

\usepackage{graphicx}	% Including figure files
\usepackage{amsmath}	% Advanced maths commands
\title[Relative Habitability with Giant Planets]{Relative Habitability of Exoplanet Systems with Two Giant Planets}

\author[Bailey \& Fabrycky]{
Nora A. Bailey,$^{1}$\thanks{E-mail: norabailey@uchicago.edu}
Daniel C. Fabrycky,$^{1}$
\\
$^{1}$Department of Astronomy \& Astrophysics, University of Chicago, Chicago, IL 60637, USA
}

\date{Accepted XXX. Received YYY; in original form ZZZ}

\pubyear{2022}

\begin{document}
\label{firstpage}
\pagerange{\pageref{firstpage}--\pageref{lastpage}}
\maketitle

\begin{abstract}
The architecture of a planetary system can influence the habitability of a planet via orbital effects, particularly in the areas of stability and eccentricity. Some of these effects are readily apparent, particularly when they occur on short timescales that are easily numerically calculable. However, the appearance and evolution of life can take place on gigayear timescales, long enough that secular effects become important. These effects are difficult to investigate, as a direct integration requires significant computational time. In this paper, we apply a semi-analytic framework in conjunction with N-body integrations and predictive techniques to determine the relative habitability for an Earth-like planet in a system with two giant companions over a multidimensional parameter space. Relative habitability quantifies the integrated habitability probability compared to a system containing only a single Earth-like planet. We find trends with mass, eccentricity, location, spacing, inclination, and alignment of the giant planets, including configurations where the system is more habitable due to the giant planets. As long as the system remains stable, a moderate eccentricity excitation of the terrestrial planet can be beneficial by increasing the outer boundary of the habitable zone through higher mean irradiance. In our simulations, the median ($\pm 1 \sigma$) habitable planet has an eccentricity of $0.11^{+0.16}_{-0.08}$, though it started circular. Low-mass, widely separated, and moderately eccentric perturbing giants can accomplish this, an ``ultra-habitable'' configuration of companions.
\end{abstract}


%%%%%%%%%%%%%%%%%%%%%%%%%%%%%%%%%%%%%%%%%%%%%%%%%%

%%%%%%%%%%%%%%%%% BODY OF PAPER %%%%%%%%%%%%%%%%%%

\section{Introduction}

Much work has been devoted to uncovering which aspects of the solar system are necessary to the phenomenon of life---particularly intelligent life---appearing on Earth. Extrapolating to a wider variety of planetary systems, such as those common in exoplanet architectures, can help us understand the ubiquity of the conditions for life in the galaxy. The interplay of factors that render a given planet habitable is exceedingly complex. This is a cross-disciplinary field of study with many potential knobs to turn. In this paper, we approach the question from a planetary dynamics angle and examine effects of the architecture of a system on the habitability of an Earth twin---that is, a planet with the same planetary properties and host star as Earth, but with varying orbital properties and companions.

Previous work has examined the possibility of an Earth surviving in the architectures of known exoplanet systems \citep{2018Agnew,2020Kokaia} and considered future target selection \citep{2020Gascon}. Here, we consider the problem from a more theoretical perspective. What properties of giant planets might be key in determining the likely habitability of a system?

Much of the previous work has focused on instabilities due to mean-motion resonances (MMRs), as these have a strong capacity to disrupt or stabilize a system architecture. MMR-driven instabilities usually develop on short timescales \citep{2009Smith,2017Obertas}, with some exceptions \citep{1982Wisdom,2015Barnes}, but given the long timescales associated with the development of life, and particularly intelligent life, the stability of the system on secular timescales becomes important as well. Secular effects from giant planets have been shown to be important in the formation of planetary systems \citep{2016Haghighipour}. Giant planets can also interact with binary companions, whether stellar or planetary, to create secular resonances relevant to the habitable zone \citep{2017Bazso, 2019Denham}.

We aim to examine which system architectures are dynamically conducive to a habitable Earth twin on both short and long timescales. Given that Earth-like planets are sensitive to the architectures of their giant companions \citep{2020Horner}, we expect that having giant planets in a system will alter the relative habitability. Indeed, in our own solar system we see that the particulars of the configuration of the gas giant planets, Jupiter and Saturn, has implications for the habitability of our Earth \citep{2015PilatLohinger}. \cite{2019Agnew} conducted a study in a similar vein, using test particle simulations to predict the stability of habitable Earths with a giant planet---here we will consider the mass of the Earth and two giant planets. Having multiple giant planets allows for a more complex interplay of secular dynamics, which can lead to habitable zone effects even from cold giant planets.

The paper is organized as follows: in Section~\ref{sec:methods} we outline the process we use to determine the relative habitability of a given architecture of two giant planets; in Section~\ref{sec:1D} we present and discuss our results of a one-dimensional analysis of two fiducial systems; in Section~\ref{sec:grid} we present and discuss our results of a multi-dimensional analysis across eight architecture parameters; the paper concludes with Section~\ref{sec:conclusion}.

\section{Methods}\label{sec:methods}

\subsection{Architecture Selection}\label{sec:initialarch}

Given the large variety of observed planetary system architectures (e.g. \citealt{2011Lissauer,2014Fabrycky,2015Winn,2021Zhu}), the potential set of architectures to evaluate is almost limitless. For the sake of completing an initial examination in this work, we will restrict ourselves to considering strictly an Earth-like planet and two giant planet companions around a Sun-like star. The range of parameter space covered is summarized in Table~\ref{tab:params}. For giant companions, we consider a mass range from 0.1 to 10~$M_{Jup}$, covering sub-Saturns to super-Jupiters.

\begin{table}
    \centering
    \caption[]{Range of parameters varied in our set of simulations.}
    \label{tab:params}
    \begin{tabular}{cccccc}
        \hline
        parameter & unit & min & max & scale & number \\
        \hline
        $m_1$ & $M_{Jup}$ & 0.1 & 10 & linear & 4 \\
        $m_2$ & $M_{Jup}$ & 0.1 & 10 & linear & 4 \\
        $a_1$ & AU & $10^{-1}$ & 10 & logarithmic* & 8 \\
        $\alpha$ &  & $10^{-0.7}$ & $10^{-0.12}$ & logarithmic & 8 \\
        $a_\oplus$ & AU & 0.6 & 1.9 & linear & 80 \\
        $e_1$ &  & 0 & $10^{-0.04}$ & logarithmic$^\dagger$ & 6 \\
        $e_2$ &  & 0 & $10^{-0.04}$ & logarithmic$^\dagger$ & 6 \\
        $\Delta \varpi$ & rad & 0 & $\pi$ & linear & 2 \\
        inclination$^\ddag$ & rad & 0 & $\pi/9$ & linear & 2 \\ 
        \hline
        \multicolumn{6}{p{\linewidth}}{*There is a break in the logarithmic scale at $a_1=10^{0.1}$, with three evenly spaced below (inclusive) and five evenly spaced above.} \\
        \multicolumn{6}{l}{$^\dagger$The lowest eccentricity, $e=10^{-2}$, was replaced with $e=0$.} \\
        \multicolumn{6}{p{\linewidth}}{$^\ddag$There are two inclination cases, one with $i_1=i_2=i_\oplus=0$ and one with nonzero inclinations assigned to the giant planets ($i_1=14.1^\circ$, $i_2=1.4^\circ$).}
    \end{tabular}
\end{table}

Our Earth analogs are placed uniformly between 0.6 and 1.9~AU to provide information about the entire extent of the habitable zone (see Section~\ref{sec:hab} for details about our habitability model). To cover giant planet configurations that are interior, exterior, or surrounding the Earth, we place the inner giant planet (planet 1) on a range between 0.1 and 10~AU and set the location of the outer giant planet (planet 2) based on a given semi-major axis ratio, $\alpha=\sfrac{a_1}{a_2}$.

Planets that are sufficiently close to one another will experience large enough perturbations to render a system unstable. The spacing at which this happens depends on many factors, but several previous studies have established a linear relationship between the logarithm of the time for a system to become unstable and the initial planetary separation in mutual Hill radii \citep{1996Chambers, 2009Smith, 2016Morrison,2017Obertas}.  We use recent results from \cite{2021Lissauer} to set a limit for our giant planets of $\beta$~=~7.15 in mutual Hill radii spacing. We apply this limit to the least massive configuration ($m_1=m_2=0.1$~$M_{Jup}$) to set the highest $\alpha$ in our parameter space at $10^{-0.12}$ ($\sim0.76$). For the most massive configuration ($m_1=m_2=10$~$M_{Jup}$), our lowest $\alpha$ of $10^{-0.7}$ ($\sim0.20$) is above the $\beta$~=~7.15 limit.

In eccentricity, we cover a range from circular to high for both giant planets, with a maximum $e=10^{-0.04} \sim 0.91$. Given that we expect more stable systems at low eccentricity, we use a logarithmic scale to provide more resolution at low eccentricity, although we replace the lowest eccentricity ($e=10^{-2}$) with the circular case. We consider two different eccentricity alignments, the perfectly aligned case ($\Delta \varpi=0$) and the antialigned case ($\Delta \varpi=\pi$).

Finally, we consider two different inclination cases. The first is the coplanar case, where all three planets have zero inclination. The second is a case including mutual inclinations by assigning each giant planet a fixed inclination ($i_1=14.1^\circ$ and $i_2=1.4^\circ$); these values were chosen randomly and used consistently for each simulation in the non-coplanar case (the longitudes of the ascending nodes, $\Omega$, are always set to 0).

For each configuration of the giant planets, four sets of mean anomalies are randomly generated uniformly between 0 and $2\pi$ and an Earth is placed at 1~AU. The stability for each of these four systems is estimated using the \texttt{SPOCK} machine-learning model \citep{2020Tamayo}. The set of mean anomalies with the highest predicted stability is then used for each of the systems with changing Earth location generated for that giant planet configuration. This method is used to reduce the likelihood of choosing a particularly unstable initial condition for the systems.

To maintain the computational load in a reasonable timeframe, we use four variations of each of the giant planet parameters $m_1$ and $m_2$, six variations of $e_1$ and $e_2$, and eight variations of $a_1$ and $\alpha$. We use 80 steps in the Earth's semi-major axis, as we calculated this to be the minimum number of steps to resolve the widths of secular resonances that fall in the habitable zone. A numerical test from a small sample of configurations using 160 steps saw no significant variation in results, confirming that we are not resolution-limited at this number. With the two $\Delta \varpi$ cases and the two inclination cases, this brings our total number of configurations evaluated to $1.179648\times10^7$, with 147,456 different giant planet configurations evaluated for relative habitability.

\subsubsection{Fiducial Systems}\label{sec:fid}

In addition to this complete multi-dimensional set of architectures, we also conduct a one-dimensional analysis of two fiducial systems varying one parameter at a time. This allows us to increase the resolution of the parameters by a factor of ten. The parameters of the first fiducial system are set at $a_1$~=~4~AU, $\alpha$~=~0.32, $m_1=m_2=1$~$M_{Jup}$, $e_1=e_2=0.05$, $i_1=i_2=i_\oplus=0$, and $\Delta \varpi=\pi$. The parameters of the second fiducial system are the same except that $a_1$~=~0.15~AU, making the second fiducial system one in which the giant planets are interior to the Earth analogs. In this one-dimensional analysis, we also vary the inclination of each planet in 50 steps from 0 to $\pi$ and the eccentricity alignment of the giant planets in 20 steps from 0 to $\pi$. The mean anomalies are randomly chosen for the fiducial system and remain fixed for all simulations.

\subsection{Process}\label{sec:process}

For each 3-planet system, we utilize the following procedure for determining the stability outcome of that configuration. If at any step the configuration is deemed unstable, the outcome is set to zero and the remaining steps are skipped.

This is an overview; the details of each part of the process are discussed further in the following sections. 

\begin{enumerate}
  \item Check if the giant planets alone are expected to be stable based on the 2-planet analytical stability criterion of \cite{2018Hadden}.
  \item \textit{If minimum period ratio $> 1.4$:} Calculate the minimum orbital separations of all planets over $10^9$ orbits using Laplace-Lagrange theory. If any are less than zero, the system is considered unstable.
  \item Calculate stability prediction from \texttt{SPOCK}'s machine-learning model. If zero, the system is considered unstable.
  \item Calculate the spectral fraction of the system as described by \cite{2020Volk}. This calculation requires integrating the system for $5\times10^6$ orbits of the inner planet. If the system goes unstable during the integration (as determined by orbit crossing, colliding with the star, or an unbound planet), the system is considered unstable. If the maximum AMD spectral fraction is above the stability threshold of 0.05, the system is considered unstable.
  \item If the system has not been ruled unstable, the stability outcome is set to the stability prediction value from \texttt{SPOCK}.
\end{enumerate}

We also calculate the maximum eccentricity of the Earth for each configuration, taken from the $5\times10^6$-orbit integration done during the spectral fraction calculation. The outcome values, outcome codes (which indicate which step, if any, flagged the system as unstable), and maximum Earth eccentricities are the outputs saved for each configuration.

Using both the stability outcome and a habitable zone model that takes into account eccentricity, we determine the probable habitability of each three-planet configuration. We then integrate this habitability probability over the range of Earth semi-major axes for each giant planet configuration and compare to the integrated area for an unperturbed system. This calculation gives our final results: a relative habitability for each giant planet configuration.

Throughout this work, we do not include error estimates on our relative habitability results. There are many sources of potential error, including: each of the instability predictions; the outcome of a given N-body simulation; the effect of the planets' mean anomalies; the resolution of our Earth semi-major axes; the resolution in time of an N-body simulation and the maximum eccentricity calculation; our choice of habitability model and the associated calculations. Without a method of quantifying these systematic uncertainties, we choose instead to omit assigning an error, and we primarily focus on relative trends more than exact numbers in our results.

\subsection{Stability Outcomes}\label{sec:stability}
\subsubsection{Instabilities from Mean-Motion Resonances}\label{sec:unstableMMR}

Instabilities in multi-planet systems are often driven by the overlap of MMRs \citep{1980Wisdom, 2008Mardling, 2013Deck}. For two-planet systems, \cite{2018Hadden} find an analytical solution that predicts the onset of instability, accounting for resonances of all orders. We use their Equation~16 to calculate the relative eccentricity, $Z$, of the two giant planets and their Equation~19 to calculate the approximate critical $Z$ for the onset of chaos, $Z_{crit}$. If $Z \geq Z_{crit}$, the system is considered unstable.

This criterion neglects the Earth analog. Given that the mass ratio between the Earth and the other planets is at most $1.57 \times 10^{-2}$, we do not expect the addition of the Earth analog to significantly alter the chaos map of the system. Additionally, throughout all our stability analysis, we consider any planet going unstable to be an unstable system. It is possible that the Earth analog could survive instability between the giant planets (i.e. \citealt{2020Kokaia}), but the resulting post-instability configuration of the system would not be the same as the configuration we were considering. Thus our giant planet configurations should be seen as a final state of the system---that which might be observable---rather than an initial state.

For three-planet systems, both 2-body and 3-body MMRs interact in a complex manner that makes analytical predictions of instability difficult. \cite{2020Tamayo} use machine-learning techniques to develop a model that can predict long-term instability (up to $10^9$ orbits) from a numerical N-body integration of only $10^4$ orbits. This model is implemented in the Stability of Planetary Orbital Configurations Klassifier (\texttt{SPOCK}) package\footnote{https://github.com/dtamayo/spock}. Because the model is trained with an emphasis on MMRs, it is primarily useful for identifying systems that are unstable due to MMRs.

\texttt{SPOCK} was trained to identify instabilities in compact multiplanet systems; that is, systems with a minimum period ratio less than 3. Over 60\% of our simulated systems are considered compact. For the remaining widely separated systems, the \texttt{SPOCK} machine learning model must generalize from the training set. \cite{2020Tamayo} showed the model performs well even outside its training set, in a set of systems similar to ours, in their section ``Generalization to Uniformly Distributed Systems.''

The stability prediction from \texttt{SPOCK} takes the form of a probability between 0 (completely unstable) and 1 (completely stable). For any system that is not considered unstable from another mechanism, we use the calculated stability prediction as the system's outcome.

\subsubsection{Secular Instabilities}\label{sec:unstablelong}

It is possible for a system to be stable to MMR-related instabilities but still develop long-term instabilities on secular timescales. These instabilities may arise due to secular chaos and overlapping secular resonances; for example, the marginal stability of Mercury within the solar system \citep{2008Laskar,2011Lithwick}.

Evaluating the secular stability of a system is a nontrivial problem. Direct numerical integrations take an enormous amount of computing power, and analytical predictions are complicated by the required high-order dependencies to capture complicated secular behavior for non-circular, non-coplanar systems (e.g. \citealt{2008Libert,2008Migaszewski,2020Bailey}). We take two approaches to evaluating secular stability.

The first is applying the Laplace-Lagrange approximation to the system to calculate the eccentricities of all three planets over $10^9$ orbits \citep[Ch. 7]{1999Murray}. This approximation suffers from losses in accuracy with increasing $e$, $i$, and/or $\alpha$. Thus we only use this in systems where the smallest period ratio is greater than 1.4. We chose this value based on testing of fifty systems comparing maximum eccentricities between Laplace-Lagrange theory and N-body results. Further, we use this analytical treatment to test for extremes and not to determine exact orbital parameters for the system (i.e., we do not consider the Laplace-Lagrange solution when setting the maximum Earth eccentricity during the habitability analysis).

We find the three points in time when each of the planets has its maximum eccentricity. Because the Earth is much less massive than the giant planets, its calculated eccentricity is less accurate. For example, if the Earth lies in a secular resonance, its eccentricity can be excited to $>1$ according to Laplace-Lagrange theory, which breaks down when approaching secular resonance. Therefore we include the Earth as a massive planet in this approximation, but when analyzing these three points we set the Earth eccentricity to zero rather than using its calculated eccentricity. For all three points in time, we calculate the orbital separations between the planets. If any separation is negative, we consider that system to be unstable due to predicted orbit crossing.

The second method we use for determining secular stability is the spectral fraction method. \cite{2020Volk} developed the technique of using a system's spectral fraction over a $5\times10^6$-orbit integration to predict stability on a timescale of $5\times10^9$ orbits. 

Spectral fraction is calculated as the fraction of frequencies in a fast Fourier transform of a time series of orbital elements that have a power of $\geq$5\% of the peak frequency's power. \cite{2020Volk} found that the angular momentum deficit (AMD) spectral fraction is the most predictive of instability, where AMD for a given planet is calculated as 

\begin{equation}
    \textup{AMD}~=~\frac{m M_\star}{m + M_\star} \sqrt{G(m+M_\star)a} \: (1-\sqrt{1-e^2}\:  \textup{cos}\,  i),
\end{equation}

with $G$ being the universal gravitational constant; $M_\star$ the mass of the star; and $m$, $a$, $e$, and $i$ the mass, semi-major axis, eccentricity, and inclination of the planet. For systems not strongly influenced by MMRs, a high spectral fraction (above $\sim$0.01-0.02) is correlated with a low chance of long-term stability. Given that we remove the MMR-unstable systems in an earlier portion of our analysis, we can use the spectral fraction to remove systems likely to be long-term unstable due to secular effects.

We use the \texttt{REBOUND} package \citep{2012Rein} with the WHFast integrator \citep{2015Rein2} (50 steps per innermost orbit) to perform the integrations for each system. The AMD for each planet is calculated over 3000 steps, and a real one-dimensional discrete Fourier transform is used to calculate the AMD power spectra.

\paragraph{Spectral Fraction Testing}

We tested the performance of the AMD spectral fraction in predicting long-term instability by creating 100 random systems consisting of an Earth and two Jupiters around a Sun-like star.

The inner Jupiter's period was randomly chosen uniformly between 3000 and 5000 days, and the period ratio with the outer Jupiter was randomly chosen in log space between 1.77 and 10, avoiding first and second-order MMRs.

Using Laplace-Lagrange secular theory, we found the predicted locations of the secular resonances of this Jupiter pair. The Earth was then placed in between the locations predicted for the inclination-node secular resonance and one of the eccentricity-pericenter secular resonances. This method was used to create configurations that might be in regions affected by secular resonance overlap. Additionally, given our interest in the dynamical habitability, it was enforced that the Earth must fall into the habitable zone, using the limits of \cite{2013Kopparapu} between 0.97 and 1.7~AU.

Mild initial eccentricities and inclinations were assigned to all three planets using Rayleigh distributions with scales of 0.049 for $e$ \citep{2015VanEylen} and 0.032 rad for $i$ \citep{2014Fabrycky}.

Lastly, we required that \texttt{SPOCK} must predict stability of at least 0.70. Our intended use of the spectral fraction calculation is to determine instability among systems not already predicted to be unstable.

With this set of 100 Earth-Jupiter-Jupiter systems, we calculated the AMD spectral fraction of the Earth as described in Section~\ref{sec:unstablelong}. We also integrated each of these systems for $5\times10^9$ orbits of the Earth using the WHFast integrator in \texttt{REBOUND}, using 50 steps per orbit.

\begin{figure}
	\centering
    \includegraphics[width=\linewidth]{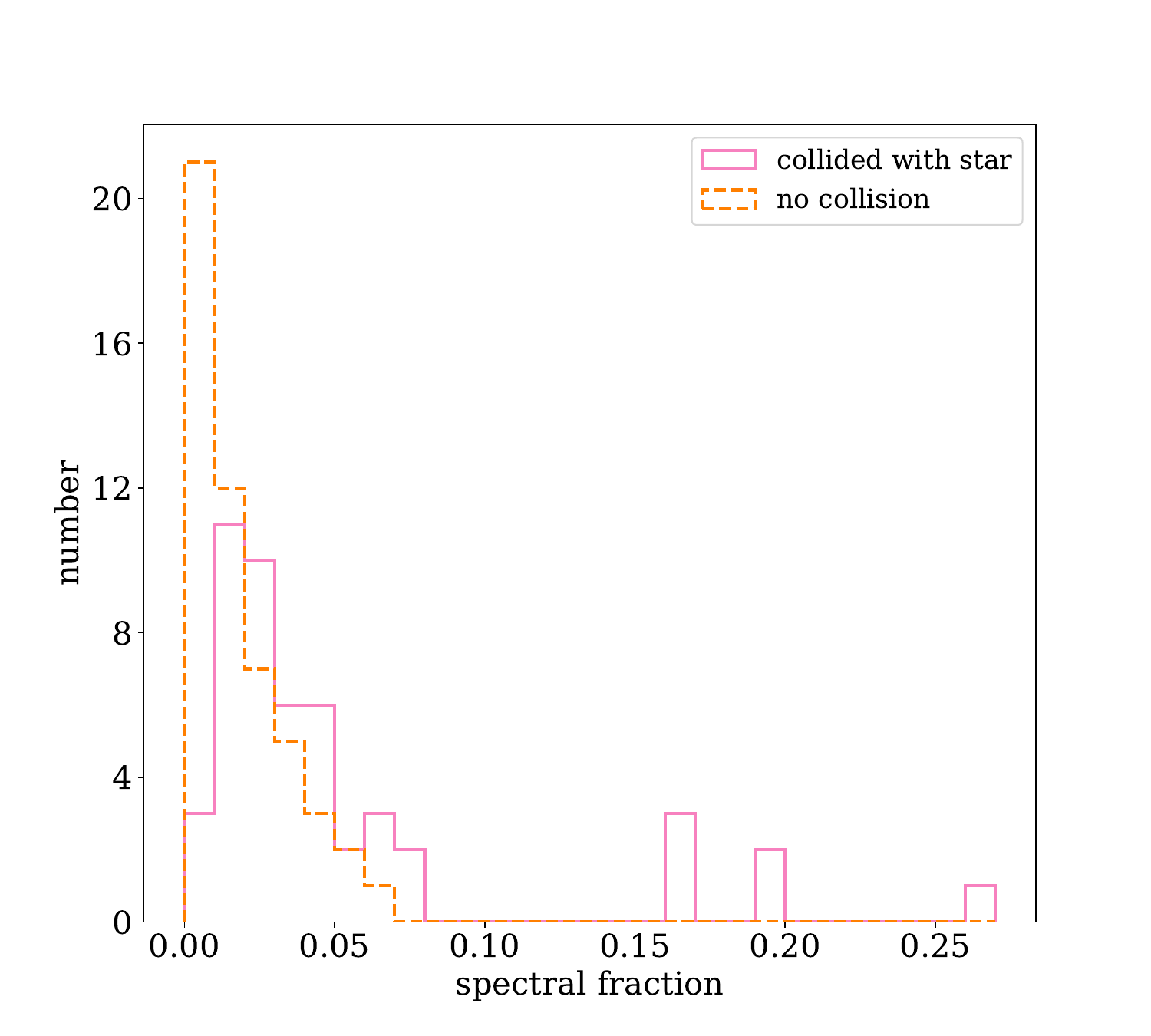}
	\caption[]{The distribution of AMD spectral fractions for the Earth in an Earth-Jupiter-Jupiter system for cases where the Earth collided with the central star and systems where it did not. A high spectral fraction is predictive of instability, but a low spectral fraction is not predictive of stability.}
	\label{fig:SF_100_hist}
\end{figure}

No system had explicit orbit crossing or unbound planets over $5\times10^9$ orbits. However, the Earth's eccentricity was excited in the vast majority of systems, including 49 in which the Earth had a pericenter passage of less than 1 stellar radius, i.e., it would have collided with the star. All of the systems with a spectral fraction greater than 0.07 resulted in a collision. The median spectral fraction for collision systems was 0.031, while the median spectral fraction for non-collision systems was 0.014. The distribution of spectral fractions is shown in Figure~\ref{fig:SF_100_hist}.

From this test, we conclude that, for systems such as we are interested in, the spectral fraction method is a good indicator of instability but not of stability. That is, a high spectral fraction meant the system was unstable, but the reverse was not true. Therefore we do not predict stability from spectral fraction, but we do use it to determine that a system is unstable. We use 0.05 as our threshold, slightly higher threshold than that of \cite{2020Volk}.

\subsection{Habitability Model}\label{sec:hab}

Habitability is an enormously complex topic with many complicating factors, many of which aren't even yet known. In order to quantify the habitability of a given system in this work, we use a probabilistic habitability model. The model we adopt is relatively simple, but this process could be adapted to work with models of varying complexity to incorporate many more aspects of habitability.

The habitable zone (HZ) is a concept referring to the region around a star in which a planet can sustain liquid water on its surface. The location of the HZ is affected by planetary factors such as the planet's size, atmosphere, water content, etc. and by stellar factors such as spectral type and temperature. Numerous studies have examined the extent of the HZ in various systems; see \cite{2020Kopparapu} for a review.

\begin{figure}
	\centering
    \includegraphics[width=\linewidth]{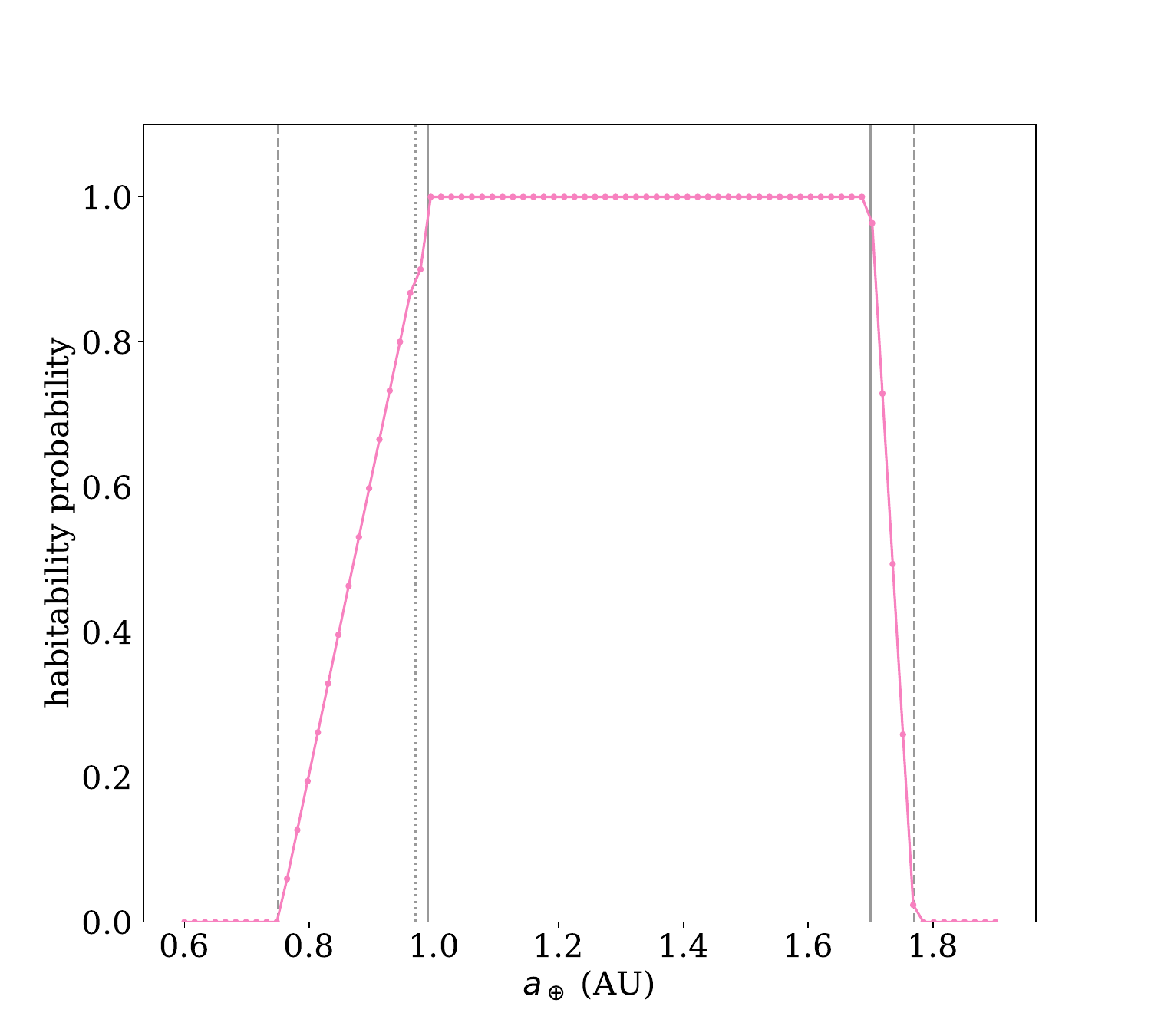}
	\caption[]{Our probabilistic model for habitability of an unperturbed Earth analog orbiting a Sun-like star. The vertical lines show the limits of the habitable zone from \cite{2013Kopparapu} (0.75 AU - recent Venus; 0.97 AU - runaway greenhouse; 0.99 AU - moist greenhouse; 1.70 AU - maximum greenhouse; 1.77 AU - early Mars). The points show the resolution in the Earth's semi-major axis used in our analyses.}
	\label{fig:unperturbed_HZ}
\end{figure}

The basis for our model is the limits of the HZ from \cite{2013Kopparapu} for circular ($e=0$) planets. Between their conservative limits of 0.99~AU and 1.7~AU (based on the moist greenhouse limit for the inner HZ boundary and the maximum greenhouse for the outer HZ boundary), we set the probability of habitability to one. Between 0.99~AU and the runaway greenhouse limit, 0.97~AU, we set the probability of habitability to 0.9. Inwards of their optimistic limit of 0.77~AU for the inner HZ, based on the early Venus limit, and outwards of their optimistic limit of 1.77~AU, based on the early Mars limit, we set the probability of habitability to zero. Between the conservative and optimistic limits (between 0.97~AU and 0.77~AU on the inner edge and between 1.7~AU and 1.77~AU on the outer edge), the habitability probability decreases linearly. See Figure~\ref{fig:unperturbed_HZ} for an illustration of this model. We take this to be our ``unperturbed'' system that we will use for comparison with our perturbed systems. We note that we are comparing only perturbations from the giant planets; it is possible that an Earth-only system has eccentricity excitation from another mechanism (disk interactions, quadrupole moment of the star, etc.), which are neglected in our ``unperturbed'' reference system.

In addition to the semi-major axis of the Earth, we also wish to consider the impact of the Earth analog's eccentricity, which could become excited by the architecture of the system once the giant planets are added. The effect of eccentricity on the location of the HZ is a complex problem that has been the subject of many studies (e.g. \citealt{2002Williams,2010Dressing,2012Kane,2015Linsenmeier, 2016Bolmont,2017Way,2017Mendez,2017Kane,2020Palubski}). A common approach is to use the mean flux approximation, which uses the average stellar flux received by a planet on an eccentric orbit to compare with the equivalent location of a circular orbit. However, \cite{2016Bolmont} use detailed climate models to show this approximation becomes unreliable at high eccentricities.

Eccentric HZ limits based on climate models, both 1-D and 3-D, are more accurate than analytical approximations but, due to extensive computation time and the profusion of variable parameter combinations, are difficult to use for generalized limits. Climate models have shown that increasing eccentricity can move the outer HZ limit further out \citep{2002Williams,2010Dressing}, though the effect on the inner HZ limit is less well-studied. Further complication arises from the fact that seasonality is strongly affected by both eccentricity and obliquity \citep{2015Linsenmeier}, but we are neglecting the obliquity in this analysis.

For our model, we use the mean flux analytical approximation for eccentricities below 0.6. Because the mean flux scales with $(1-e^2)^{-1/2}$ \citep{2002Williams} for a given semi-major axis, and flux also scales with the inverse square of the semi-major axis, we can calculate an equivalent circular semi-major axis that receives the same mean flux as the planet with semi-major axis $a$ and eccentricity $e$:

\begin{equation}
    a_{\left< F \right>}~=~a (1-e^2)^{1/4}
\end{equation}

For higher eccentricities, where the mean flux approximation has been shown to be invalid, we look to \cite{2020Palubski}. They use a 1-D energy balance model to evaluate the eccentric habitable zone for Earth-like planets. Their Figure 7 shows how the habitable zone varies for different eccentricities and insolation. Their results shows a steeper relation than the $(1-e^2)^{1/4}$ scaling from the mean flux approximation. We fit by eye a scaling exponent of $(1-e^2)^{1/2.1}$ and use this scaling for planets with eccentricities greater than 0.6.

\begin{figure}
	\centering
    \includegraphics[width=\linewidth]{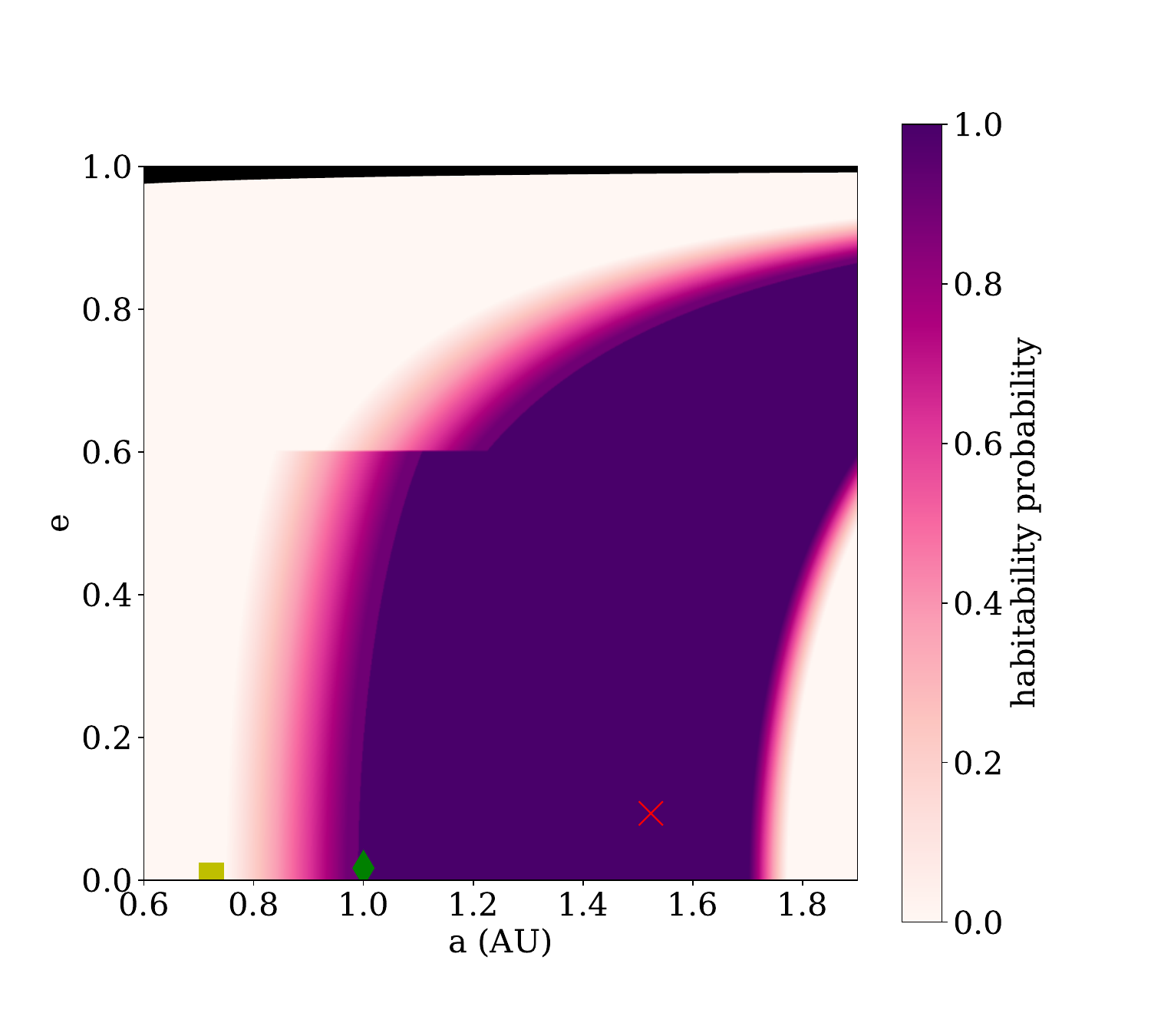}
	\caption[]{Our probabilistic model for habitability of an eccentric Earth analog orbiting a Sun-like star. The upper black region is disallowed due to collision with the star; present-day Venus (yellow square), Earth (green diamond), and Mars (red X) are shown for reference. Probabilities are calculated based on the circular HZ limits of \cite{2013Kopparapu} and the eccentricity scaling of the mean flux approximation for $e<0.6$ and an estimated scaling based on \cite{2020Palubski} for $e\geq0.6$.}
	\label{fig:ae_HZ}
\end{figure}

Figure~\ref{fig:ae_HZ} shows the habitability probability calculated by our model in $e-a$ space. Note the discontinuity at $e=0.6$ due to our dual-method approach. The semi-major axis range shown is truncated based on where we place our Earth analogs; the outer boundary of the HZ continues to move outwards for increasing eccentricity, but the area of parameter space grows smaller and the likelihood of stability at these high eccentricities is low.

To determine the habitability probability for a given Earth analog, we calculate its equivalent circular semi-major axis based on its eccentricity via this model and then calculate the habitability probability as described above for circular planets.

\subsection{Example Relative Habitability Calculation}

The output that we achieve from our combination of stability and habitability models is a single number that quantifies the relative habitability for a given configuration of two giant planets. In this section, we will illustrate how these models are used to provide our calculated output for one particular giant configuration.

This giant planet configuration is taken from the architectures based on the first fiducial system with $a_1$~=~4~AU, $\alpha$~=~0.32, $m_1=1$~$M_{Jup}$, $e_1=e_2=0.05$, $i_1=i_2=i_\oplus=0$, and $\Delta \varpi=\pi$, and here we take the mass of the outer giant planet ($m_2$) to be 10~$M_{Jup}$.

\begin{figure}
	\centering
    \includegraphics[width=\linewidth]{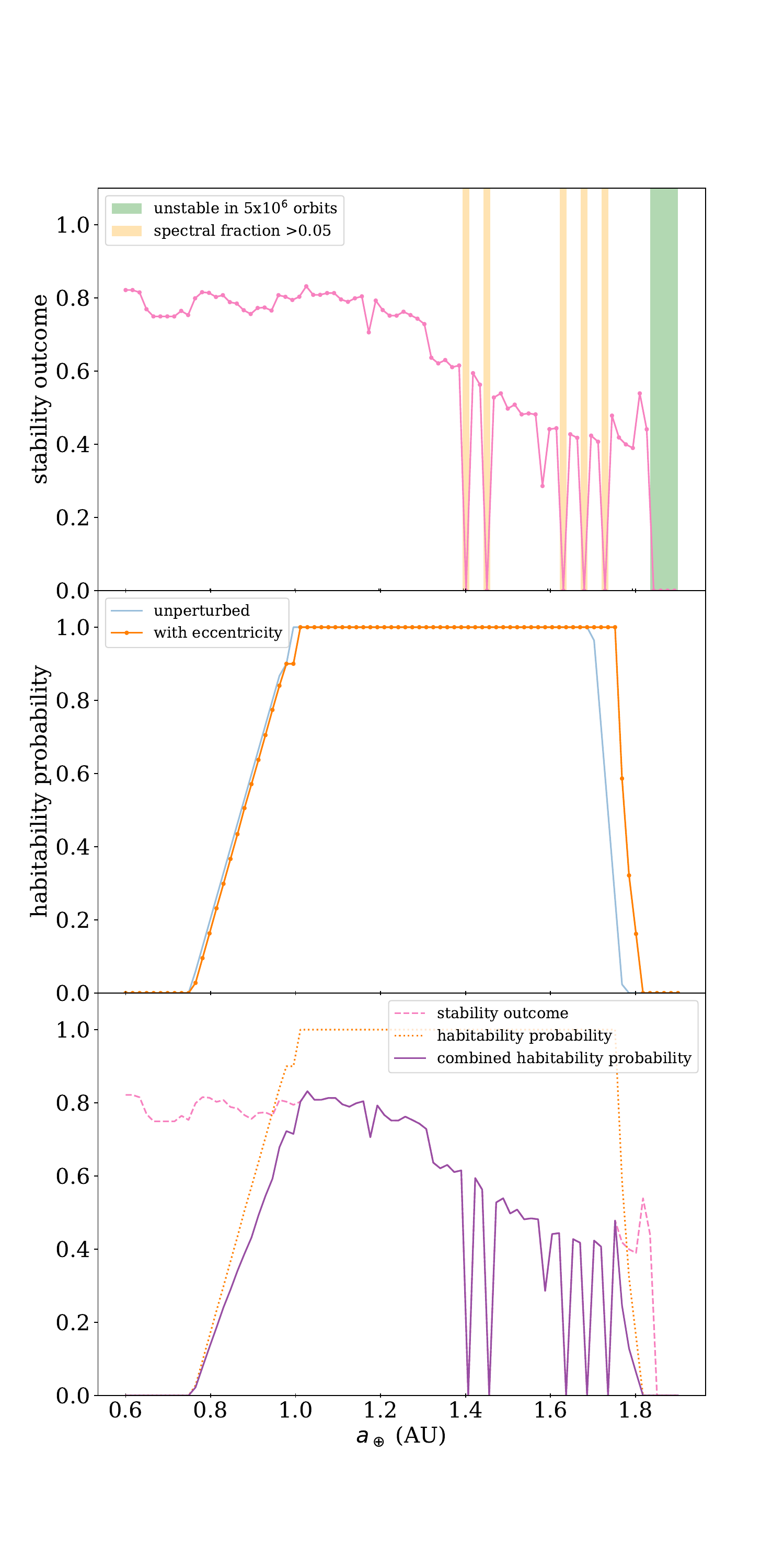}
	\caption[]{Illustration of the calculation of relative habitability for a single giant planet configuration. (top) The stability outcome for each of the 80 Earth locations. (middle) The habitability probability for each Earth given its location and eccentricity. The unperturbed habitability from Figure~\ref{fig:unperturbed_HZ} is shown for reference. (bottom) The combined relative habitability curve obtained by multiplying the stability outcome and habitability probability. Integrating the area under the combined relative habitability curve and comparing with the area under the unperturbed habitability curve (Figure~\ref{fig:unperturbed_HZ}) gives the relative habitability for a given giant planet configuration (0.6152 for this example).}
	\label{fig:example_relhab_calc}
\end{figure}

First, we get stability outcomes for each of the 80 Earth semi-major axes. The stability outcomes are shown in Figure~\ref{fig:example_relhab_calc} (top). The non-zero stability outcomes all come from the the \texttt{SPOCK} prediction. There are many reasons to get a stability outcome of zero (unstable), and the shaded background color in Figure~\ref{fig:example_relhab_calc} (top) indicates which step of the process (as described in Section~\ref{sec:process}) resulted in the unstable determination. This information is used only for our understanding, as the stability outcomes are treated the same regardless of their origin. In this example, we see instability during the $5\times10^6$-orbit numerical integration and instability predicted from a high spectral fraction.

Second, we calculate the habitability probability for an Earth analog at each given semi-major axis and eccentricity. We use the maximum eccentricity reached during the $5\times10^6$-orbit numerical integration for this; for any configuration that did not reach this step due to instability, the eccentricity remains at zero---this has no effect on the final output as the stability outcome is zero for those configurations. The variation in eccentricity over time may have additional habitability implications \citep{2017Way}, which we neglect here. Because our Earths may sometimes reside outside the outer boundary of the HZ, when developing our model we used the ``cold start'' results from \cite{2020Palubski}, which consider planets initially frozen. 

The habitability probability for this example is shown in Figure~\ref{fig:example_relhab_calc} (middle). The overall shape is almost the same as the unperturbed habitability in Figure~\ref{fig:unperturbed_HZ}; this is common for most of our results as the eccentricity effect tends to be quite small. In this case, however, there is an increased habitability probability in the outer edge of the HZ that arises from the Earth getting an excited eccentricity that makes it more habitable in our model.

We combine our stability outcome and habitability probability into a combined habitability probability by multiplying them together for each Earth semi-major axis. The resulting curve in this case is shown in Figure~\ref{fig:example_relhab_calc} (bottom). At most points, either the stability outcome or the habitability probability dominates but, especially near the edges of the HZ, they both contribute to the result.

Finally, we integrate the area underneath this combined habitability probability curve. This is done numerically using an unsmoothed univariate spline as implemented in \texttt{SciPy} \citep{SciPyref}. Here, the resulting area is 0.5304. We divide this area by the area underneath the unperturbed habitability curve (0.8649) to obtain the relative habitability for this giant planet configuration, given a uniform distribution of Earth-like planets: 0.6152.

\section{One-Dimensional Analysis}\label{sec:1D}

\subsection{Giants Exterior to HZ}\label{sec:1Douter}
\subsubsection{Results}\label{sec:1Douterresults}

The first fiducial system we consider has a semi-major axis of 4~AU for the inner giant, placing the giants exterior to the HZ (except when $a_1$ is the parameter varied). See Table~\ref{tab:params} and Section~\ref{sec:fid} for details.

\begin{figure*}
	\centering
    \includegraphics[width=\linewidth]{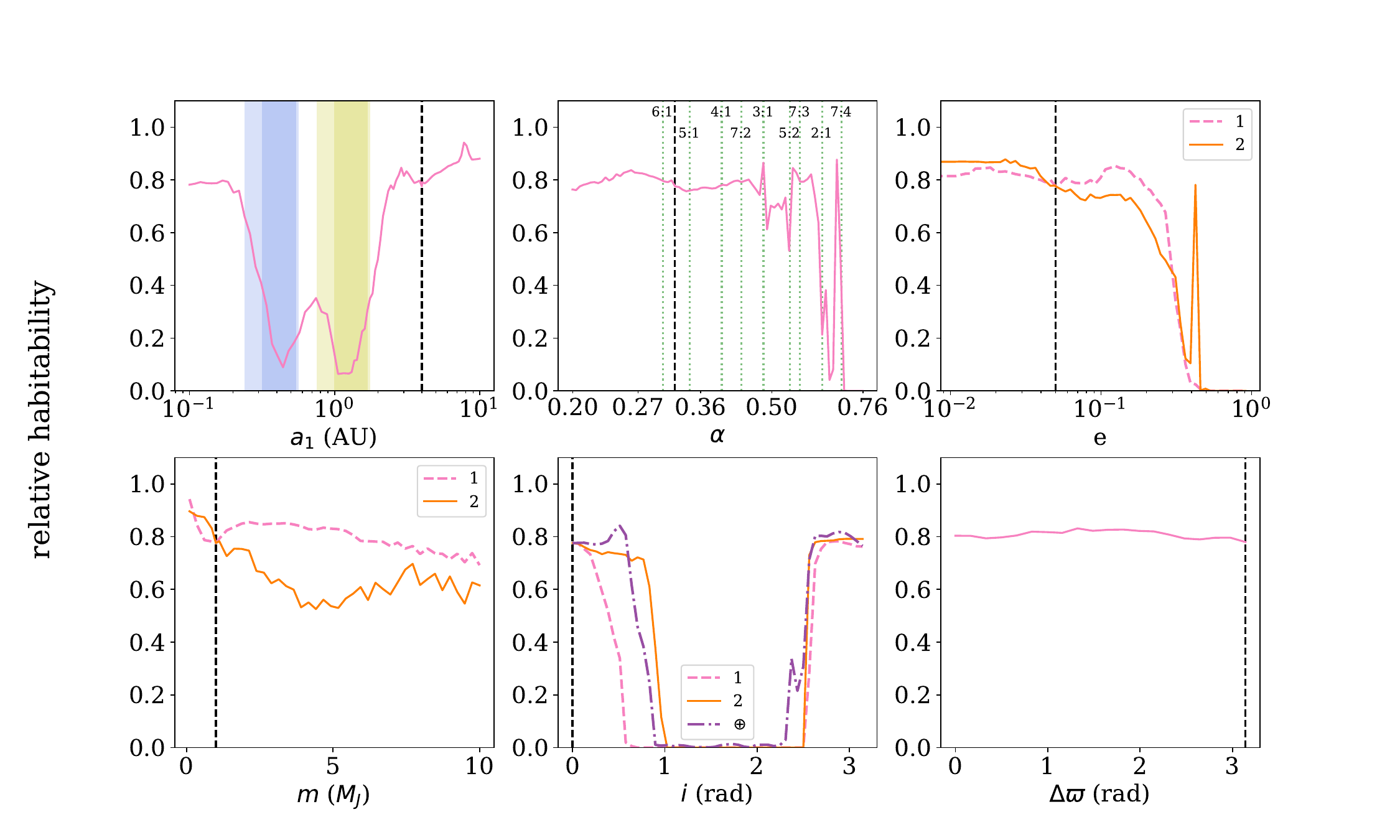}
	\caption[]{Relative habitability calculated for varied parameters by changing a single parameter of a fiducial system with exterior giant planets. The values of the fiducial system are indicated in each plot by a dashed vertical line. In the top left plot ($a_1$), the shaded background indicates when one of the giant planets is located within the habitable zone (yellow for the inner giant and blue for the outer giant; darker for the conservative limits and lighter for the optimistic limits). In the top center plot ($\alpha$), several low-order mean-motion resonances have been indicated by dotted green lines and labeled with the associated resonance.}
	\label{fig:outerfidresults}
\end{figure*}

The resulting relative habitability curves for each parameter are shown in Figure~\ref{fig:outerfidresults}. Parameters that have the same range ($m_1/m_2$, $e_1/e_2$, and $i_1/i_2/i_\oplus$) are shown on the same axes for compactness, but all the parameters were varied independently. The values of the fiducial system are indicated on each plot.

\subsubsection{Discussion}\label{sec:1Douterdisc}

There are several prominent features in Figure~\ref{fig:outerfidresults}.

First, let's consider the impact of changing the $a_1$ parameter. Note that because $\alpha$ is fixed, this also changes the location of planet 2 ($a_2$) as well. Unsurprisingly, the relative habitability decreases significantly when the giant planets are located in or near the HZ. There is a local maxima at $a_1$~=~0.745 ($a_2$~=~2.328). This is due to the small region of stability between the two giant planets falling in the middle of the HZ. We note there appear to be small peaks at $\sim$3 and 7.8~AU, but we find no clear reason that seems to explain these; it is possible they may not be significant.

As $\alpha$ increases, the giant planets get closer together by moving planet 2 inward. At $\alpha$~=~0.7 and higher, the giant planets are so close together that they are unstable by the analytical criterion of \cite{2018Hadden}. Most of the other features correspond with period ratios near commensurability, giving rise to instabilities based on MMR. For example, the first-order 2:1 MMR corresponds with an $\alpha$ of 0.63 (by Kepler's Third Law, $\alpha \sim (\sfrac{P_2}{P_1})^{-2/3}$), where there is a deep and broad dip in the relative habitability. Smaller dips occur near higher-order MMRs, such as the 5:2 resonance ($\alpha \sim 0.54$) and the 3:1 resonance ($\alpha \sim 0.48$).

At high eccentricities, we expect instabilities to develop as orbits cross. Nominal orbit crossing between the giant planets occurs at $e_2$~=~0.664 and between planet 1 and the outer HZ at $e_1$~=~0.558. In actuality, we see the relative habitability fall off due to instability at an eccentricity of approximately 0.3 for either giant planet. Interestingly, there is a strong spike in relative habitability at $e_2$~=~0.424. This is due to a resonance in the precession of the eccentricity vectors of the giant planets. The resonance stabilizes the system by preventing the Earth's eccentricity from being excited, which is what is causing the instability leading to low relative habitability on either side of the resonance.

As the mass of either giant planet changes, we don't see drastic peaks or dips in the relative habitability. However, there is a trend towards reduced relative habitability as the mass increases, and the trend is stronger when it is planet 2 with more mass. This seems to be a combination of two effects: there are consistently slightly more Earth locations that result in instability when a given mass is assigned to $m_2$ than $m_1$, and those instabilities tend to fall in the middle of the HZ. This trend is likely associated with the secular resonances within the systems. In Laplace-Lagrange theory, the resonance location for a test particle varies with the mass ratio of the two massive planets ($\sfrac{m_1}{m_2}$), and at mass ratios of less than about 0.5, the secular resonance locations for one of the eccentricity-pericenter resonances and the inclination-node resonance start to fall in the HZ for the fiducial configuration. The two resonance locations also become closer to one another as the mass ratio decreases, increasing the potential for overlapping resonances and secular chaos.

The same general trend is seen for all three planets when their inclination is changed. For relatively coplanar systems, prograde or retrograde, the relative habitability is largely unaffected. When one of the planets is strongly misaligned, the systems tend to be unstable. Because the Earth is interior to the giant planets, its instability arises from high eccentricities driven by Kozai-Lidov cycles once its inclination is high enough relative to the outer planets \citep{1962Kozai,1962Lidov,2011LithwickNaoz}. For misaligned prograde orbits, the relative habitability is more sensitive to the inclination of planet 1 and drops off before the critical inclination for Kozai-Lidov is reached. Even mild amounts of mutual inclination can lead to chaotic eccentricities \citep{2012Boue} that can destabilize systems.

Lastly, we see almost no effect from changing the alignment of the giant planets' pericenters. This is not unexpected, as the fiducial system has very low amounts of eccentricity. This parameter would likely have more effect for systems of higher eccentricity; for example, one can imagine the spike in $e_2$ might change in size or location for a different alignment of the planets.

\subsection{Giants Interior to HZ}\label{sec:1Dinner}
\subsubsection{Results}\label{sec:1Dinnerresults}

The second fiducial system we consider has a semi-major axis of 0.15~AU for the inner giant, placing both of the giants interior to the HZ (except when $a_1$ is the parameter varied). See Table~\ref{tab:params} and Section~\ref{sec:fid} for details.

\begin{figure*}
	\centering
    \includegraphics[width=\linewidth]{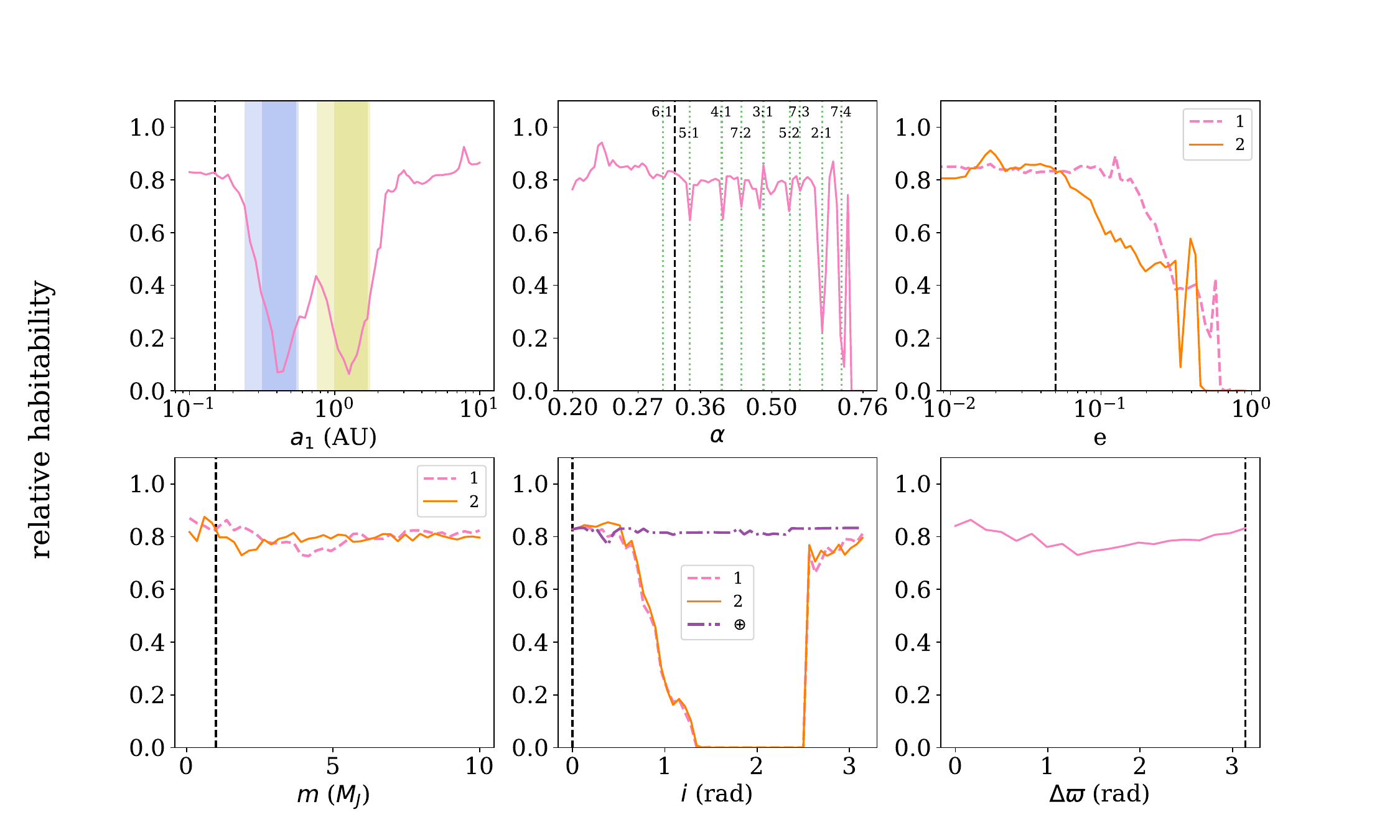}
	\caption[]{Relative habitability calculated for varied parameters by changing a single parameter of a fiducial system with interior giant planets. The values of the fiducial system are indicated in each plot by a dashed vertical line. In the top left plot ($a_1$), the shaded background indicates when one of the giant planets is located within the habitable zone (yellow for the inner giant and blue for the outer giant; darker for the conservative limits and lighter for the optimistic limits). In the top center plot ($\alpha$), several low-order mean-motion resonances have been indicated by dotted green lines and labeled with the associated resonance.}
	\label{fig:innerfidresults}
\end{figure*}

The resulting relative habitability curves for each parameter are shown in Figure~\ref{fig:innerfidresults}. Parameters that have the same range ($m_1/m_2$, $e_1/e_2$, and $i_1/i_2/i_\oplus$) are shown on the same axes for compactness, but all the parameters were varied independently. The values of the fiducial system are indicated on each plot.

\subsubsection{Discussion}\label{sec:1Dinnerdisc}

The relative habitability curves for the parameters of the interior fiducial system (Figure~\ref{fig:innerfidresults}) are similar to those of the exterior fiducial system, though some key differences exist.

First, the results for the $a_1$ parameter are almost identical, as expected, because these systems are the same in each case (as the second fiducial system is the same as the first with a different $a_1$). However, these were integrated and calculated independently, which provides a good check on the repeatability of our results.

For the $\alpha$ parameter, the overall behavior is quite similar to the previous results, falling off at high $\alpha$, but we see more narrow peaks emerge. These peaks also correspond with MMRs, those of even higher orders; for example, the 7:3 resonance ($\alpha \sim 0.57$), the 5:1 resonance ($\alpha \sim 0.34$), and even the 7:2 resonance ($\alpha \sim 0.44$). Interestingly, there seems to be some protection at the exact 3:1 resonance. The stronger effect from the higher-order MMRs is likely because the overall system is more compact and more easily destabilized.

For the eccentricities of the giant planets, again the trends are similar to the exterior case. The relative habitability begins to fall off at around the same $e_1$, but here the slope of the decline is less steep and some relative habitability remains out to much higher $e_1$. This can be understood as planet 1 is no longer the planet neighboring the HZ, so at high $e_1$ there are fewer opportunities to destabilize the system. For planet 2, in this case the relative habitability begins to decrease at a lower $e_2$; again, this is due to the ordering of the planets, as planet 2 is now the planet adjacent to the HZ ($a_2=0.4687$~AU). The same feature at $e_2 \sim 0.4$ appears, although with slightly less amplitude.

The trends of relative habitability with the mass of either giant planet are different here than in the exterior case. Here, it does not matter which planet's mass is changed, and there is very little decrease of relative habitability with increasing mass. Again we can look to Laplace-Lagrange secular theory to make sense of this. While the locations of the inclination-node and first eccentricity-pericenter secular resonances still enter the HZ for low mass ratios, they stay only in the inner edge of the HZ where the relative habitability is already low, and the secular resonance locations are not as close together, giving less opportunity for overlap. The second eccentricity-pericenter secular resonance does fall in the middle of the HZ, but it does so for all mass ratios and is well-separated from the other two secular resonances.  

We see a very notable difference with the effect of inclination in this case. There is virtually no effect from changing to inclination of the Earth. This is because the Earth, which has a tiny mass relative to the giants, is now the outer planet and, while it may be subject to inclination oscillations from the Kozai-Lidov effect, the eccentricity is not expected to be excited to extremes \citep{2017Naoz}. Nor is the effect of the Earth's inclination significant enough to destabilize the inner giant planets, regardless of its mutual inclination. However, we see that at high inclinations of either giant planet, there is no relative habitability. This is because of Kozai-Lidov effects on the inner planet when there is a large misalignment with the middle planet (the outer giant). We note that the effect is virtually identical no matter which planet's inclination is changed, unlike in the previous case, where the relative habitability was more sensitive to planet 1's inclination. 
Lastly, we similarly see very little effect from the relative alignments of the giant planets' pericenters. There is a small decrease near perpendicular alignments, which is likely because the giant planets are closer together in this case and so even at the same mild eccentricities there are slightly stronger dynamical effects.

\section{Multi-Dimensional Analysis}\label{sec:grid}

\subsection{Results}\label{sec:gridresults}

The compiled results for all 147,456 giant planet configurations are included in Table~\ref{tab:grid_results} (see Appendix~\ref{app:stability_results} for the stability outcome results prior to the habitability model being applied). Only a subset of the results are shown here, but the entire table is available to download\footnote{http://dx.doi.org/10.5281/zenodo.6324216}.

\begin{table*}
    \centering
    \caption[]{Relative habitability results for a multidimensional set of giant planet parameters.}
    \label{tab:grid_results}
    \resizebox{\textwidth}{!}{%
    \begin{tabular}{ccccccccc}
    \hline
    $m_1$ & $m_2$ & $a_1$ & $\alpha$ & $e_1$ & $e_2$ & $\Delta i$ & $\Delta \varpi$ & relative habitability \\
    ($M_{Jup}$) & ($M_{Jup}$) & (AU) &  &  &  & (rad) & (rad) &  \\ \hline
    0.1 & 0.1 & 1.90546071796325 & 0.19952623149689 & 0.02466039337234 & 0.91201083935591 & -0.22165681500328 & 3.14159265358979 & 0.00000000000000 \\
    0.1 & 0.1 & 6.60693448007596 & 0.19952623149689 & 0.14996848355024 & 0.06081350012787 & 0.00000000000000 & 3.14159265358979 & 0.92961440310821 \\
    0.1 & 0.1 & 6.60693448007596 & 0.42798502294486 & 0.00000000000000 & 0.00000000000000 & -0.22165681500328 & 0.00000000000000 & 0.93437618202497 \\
    0.1 & 0.1 & 10.00000000000000 & 0.29222292648815 & 0.14996848355024 & 0.91201083935591 & 0.00000000000000 & 0.00000000000000 & 0.00000000000000 \\
    0.1 & 0.1 & 10.00000000000000 & 0.29222292648815 & 0.36982817978027 & 0.00000000000000 & 0.00000000000000 & 0.00000000000000 & 1.10954892015974 \\
    0.1 & 6.7 & 1.25892541179417 & 0.29222292648815 & 0.14996848355024 & 0.14996848355024 & 0.00000000000000 & 3.14159265358979 & 0.00000000000000 \\
    0.1 & 6.7 & 1.90546071796325 & 0.29222292648815 & 0.91201083935591 & 0.14996848355024 & 0.00000000000000 & 3.14159265358979 & 0.00000000000000 \\
    0.1 & 6.7 & 1.90546071796325 & 0.42798502294486 & 0.02466039337234 & 0.00000000000000 & 0.00000000000000 & 3.14159265358979 & 0.41636386063234 \\
    0.1 & 6.7 & 2.88403150312661 & 0.29222292648815 & 0.06081350012787 & 0.00000000000000 & 0.00000000000000 & 0.00000000000000 & 0.72403698437208 \\
    3.4 & 0.1 & 4.36515832240166 & 0.62682001739704 & 0.06081350012787 & 0.14996848355024 & 0.00000000000000 & 0.00000000000000 & 0.64238994491122 \\
    3.4 & 3.4 & 0.10000000000000 & 0.19952623149689 & 0.14996848355024 & 0.02466039337234 & -0.22165681500328 & 3.14159265358979 & 0.75596936971053 \\
    3.4 & 6.7 & 1.25892541179417 & 0.24146664216652 & 0.91201083935591 & 0.91201083935591 & 0.00000000000000 & 3.14159265358979 & 0.00000000000000 \\
    3.4 & 10 & 1.90546071796325 & 0.62682001739704 & 0.02466039337234 & 0.06081350012787 & -0.22165681500328 & 0.00000000000000 & 0.00000000000000 \\
    6.7 & 3.4 & 0.35481338923358 & 0.29222292648815 & 0.06081350012787 & 0.02466039337234 & 0.00000000000000 & 0.00000000000000 & 0.00488707023169 \\
    6.7 & 3.4 & 4.36515832240166 & 0.29222292648815 & 0.06081350012787 & 0.06081350012787 & -0.22165681500328 & 3.14159265358979 & 0.67268960164690 \\
    6.7 & 3.4 & 6.60693448007596 & 0.35364818096244 & 0.06081350012787 & 0.91201083935591 & 0.00000000000000 & 3.14159265358979 & 0.00000000000000 \\
    6.7 & 6.7 & 1.25892541179417 & 0.62682001739704 & 0.14996848355024 & 0.00000000000000 & 0.00000000000000 & 3.14159265358979 & 0.00000000000000 \\
    6.7 & 6.7 & 1.90546071796325 & 0.24146664216652 & 0.91201083935591 & 0.36982817978027 & -0.22165681500328 & 0.00000000000000 & 0.00000000000000 \\
    6.7 & 6.7 & 2.88403150312661 & 0.42798502294486 & 0.06081350012787 & 0.91201083935591 & -0.22165681500328 & 3.14159265358979 & 0.00000000000000 \\
    10 & 10 & 2.88403150312661 & 0.35364818096244 & 0.00000000000000 & 0.02466039337234 & -0.22165681500328 & 0.00000000000000 & 0.45969948922164 \\
    \hline
    \multicolumn{9}{c}{Note: The full version of Table~\ref{tab:grid_results} is available for download at http://dx.doi.org/10.5281/zenodo.6324216 \citep{datasets}.} \\
    \multicolumn{9}{c}{A random subset of rows are shown here for guidance regarding its form and content.}
    \end{tabular}%
    }
\end{table*}

\begin{figure*}
	\centering
    \includegraphics[width=.95\linewidth]{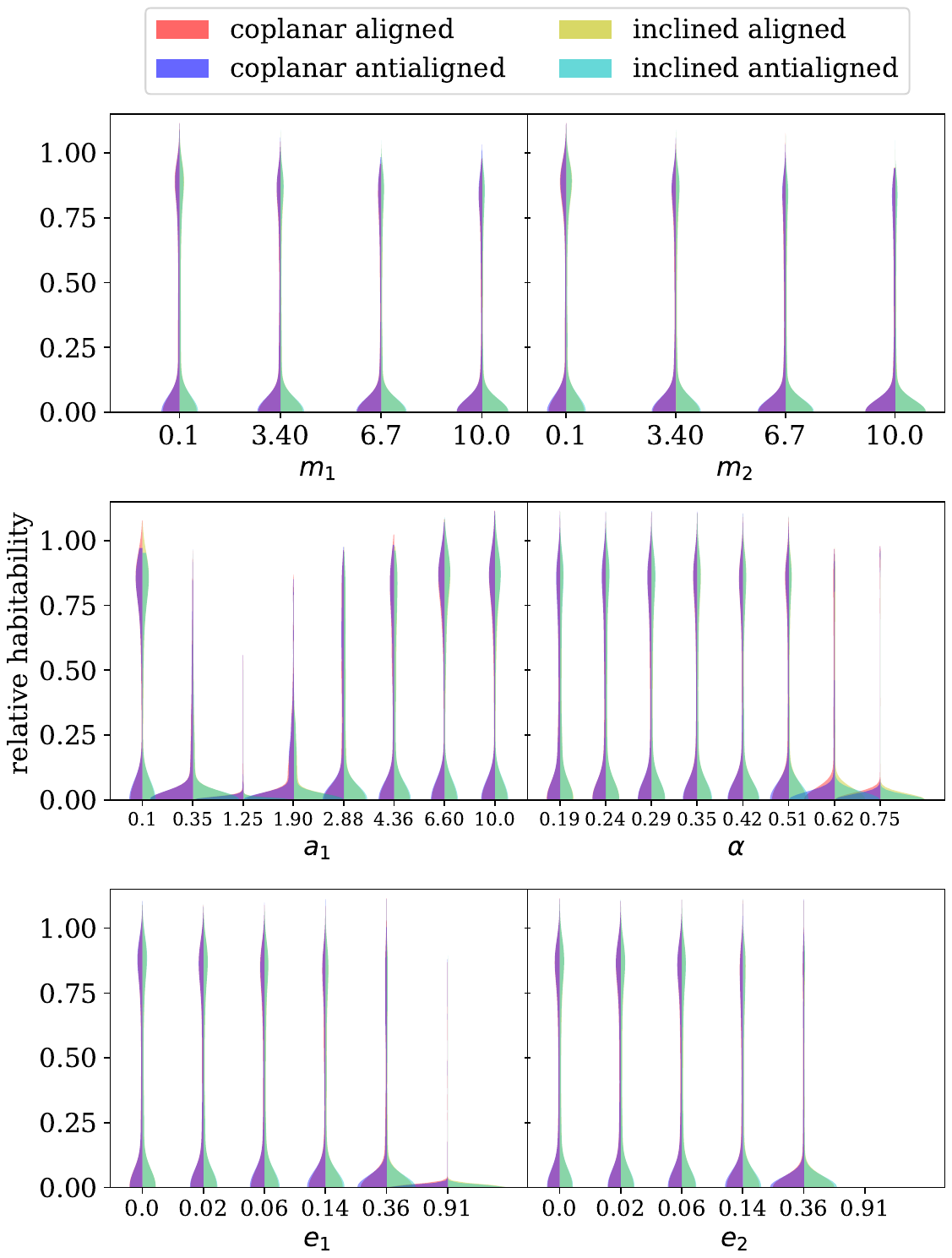}
	\caption[]{The distribution of relative habitability for each parameter. The coplanar results are on the left side of each violin plot and the inclined results on the right, with colors differentiating the aligned (red/yellow) and antialigned (blue/cyan) cases. Where the distributions overlap, the color appears purple or green. There is little visible difference between the four combinations at this level, though some small coloring variations are noted at low $a_1$ and high $\alpha$.}
	\label{fig:1D_violin}
\end{figure*}

Visualization of a high-dimensional function like these results is unfortunately limited by the few spatial dimensions we have at our disposal. To summarize these results, we show the distribution of results across each parameter in Figure~\ref{fig:1D_violin}. We show the coplanar results on the left side of each violin plot and the inclined results on the right side, with different colors for the aligned and antialigned results. These results are flattened into one dimension; that is, each distribution includes all of the results for a given value of the designated parameter. 

\begin{figure}
	\centering
    \includegraphics[width=\linewidth]{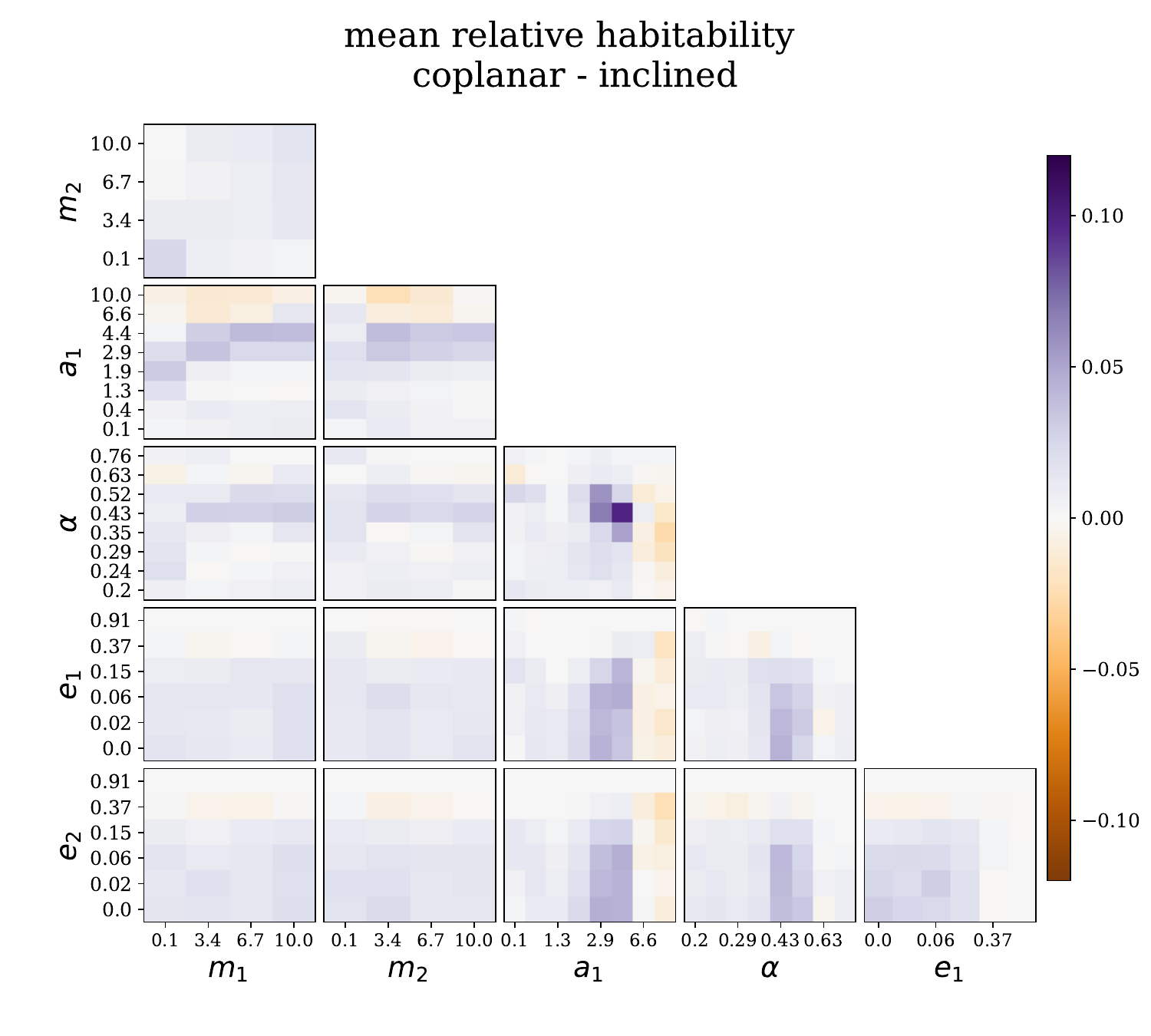}
	\caption[]{The difference in the mean relative habitability for inclined systems and coplanar systems. Purple indicates more relative habitability for the coplanar systems, while orange indicates more relative habitability for the inclined systems. The differences are mostly quite small; however, features stick out at high $a_1$ and at $\alpha~=~0.43$.}
	\label{fig:2D_incl}
\end{figure}

\begin{figure}
	\centering
    \includegraphics[width=\linewidth]{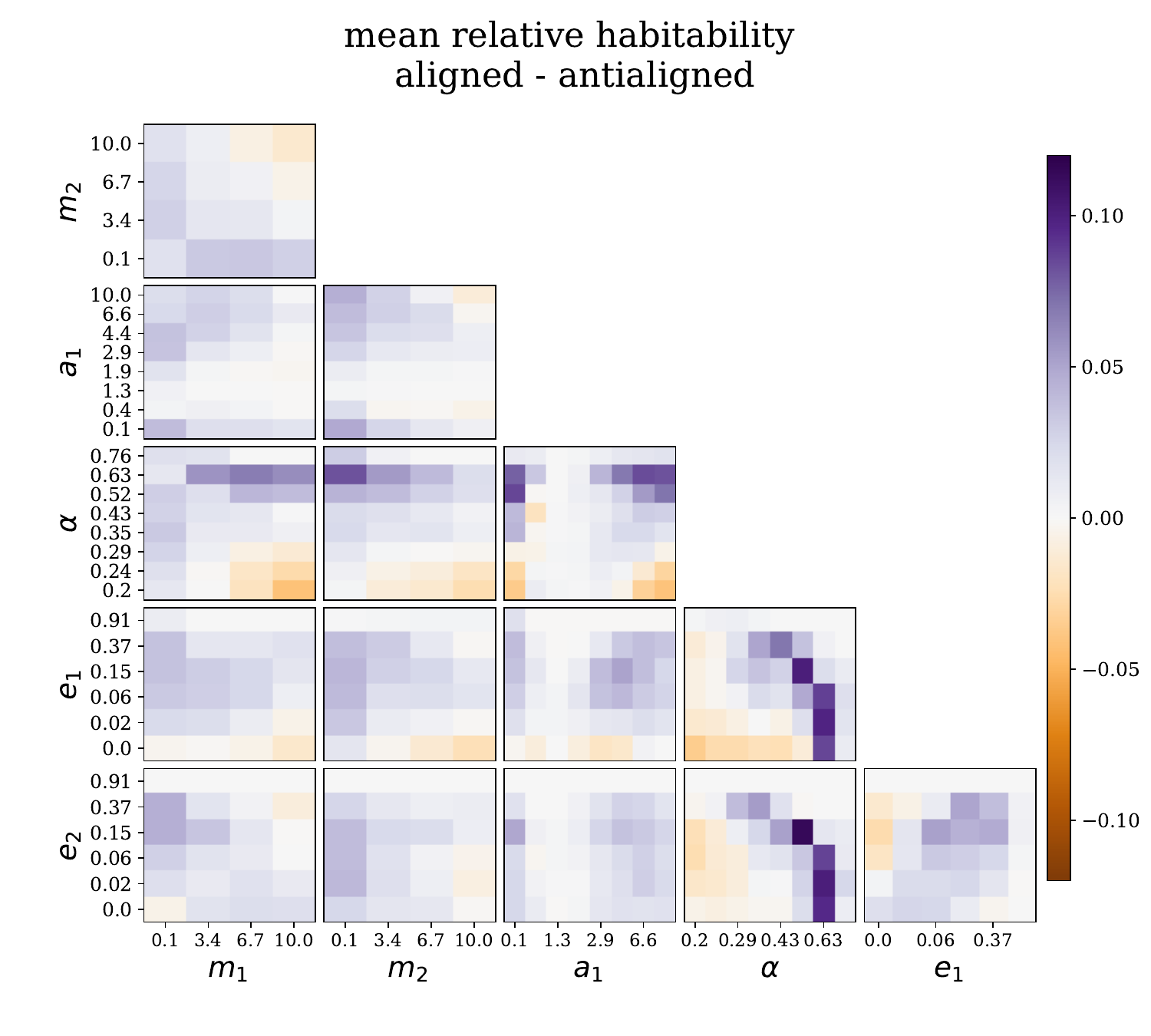}
	\caption[]{The difference in the mean relative habitability for aligned systems and antialigned systems. Purple indicates more relative habitability for the aligned systems, while orange indicates more relative habitability for the antialigned systems. The differences are mostly quite small; however, trends with $\alpha$ and $m_{1,2}$ and features at $e_{1,2}~=~0.15$ stick out.}
	\label{fig:2D_align}
\end{figure}

We note that the difference between the four combinations of inclination and pericenter alignment is very small. There appears to be some small increase in relative habitability for aligned systems over antialigned systems at very low $a_1$ and high $\alpha$. To investigate these small differences, we use a 2D differential analysis. In Figures~\ref{fig:2D_incl} and \ref{fig:2D_align}, we plot the difference in the mean relative habitability for the 2D distributions for different combinations of parameters. By using the mean, we are now flattening the relative habitability distribution into a single value, but we are reducing the flattening in parameter space by one dimension.

We've chosen to show here the comparison between the coplanar and inclined case for all systems and between the aligned and antialigned systems for all systems, as the alignment does not seem to have strong effect on the general inclination results nor vice versa.

\begin{figure}
	\centering
    \includegraphics[width=\linewidth]{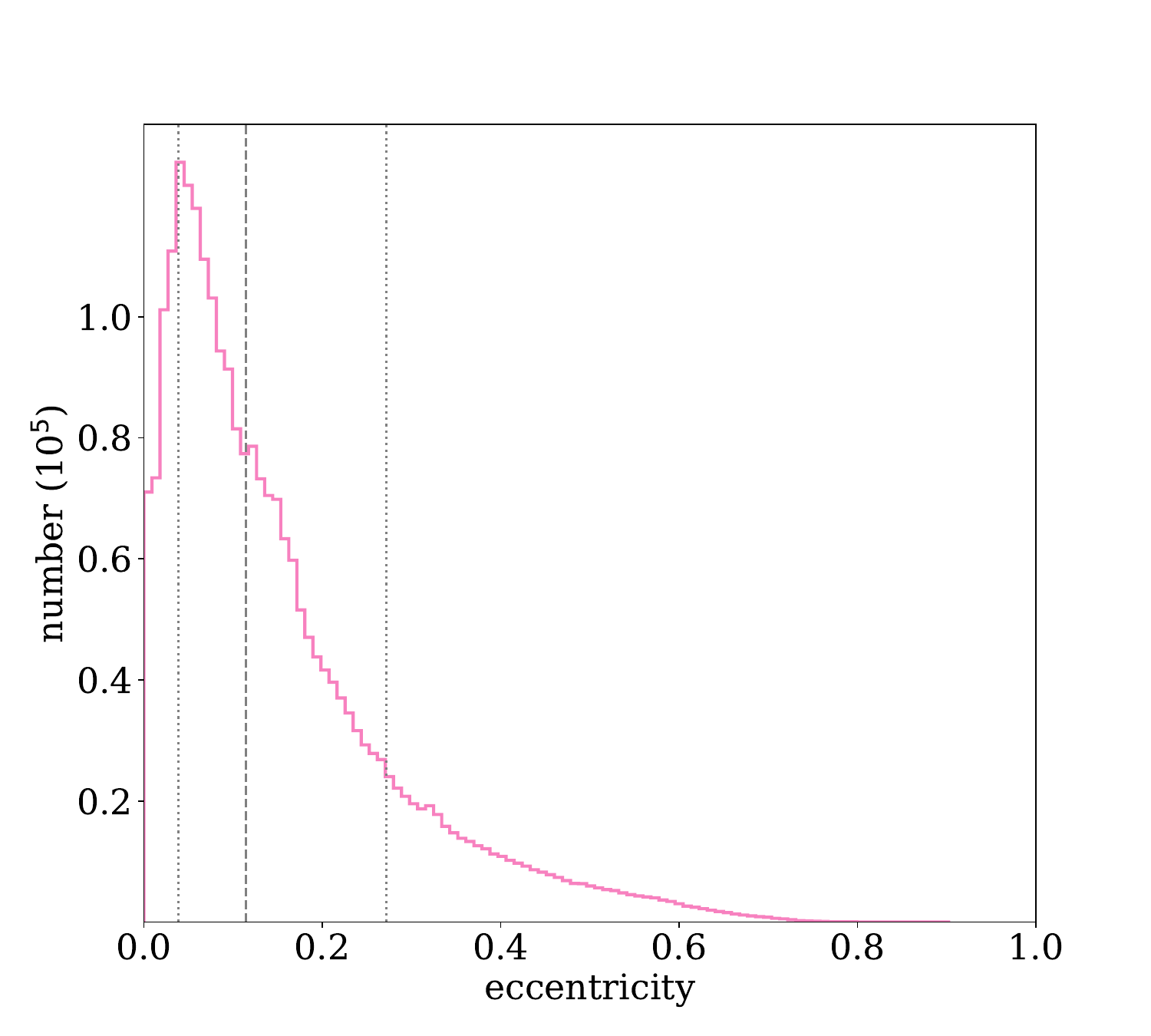}
	\caption[]{A histogram of the maximum Earth eccentricities for Earths with nonzero habitability probability. The median (0.114) and 15.9/84.1 percentiles (0.039/0.272) are indicated with vertical lines. The distribution contains two features: an excess of low-eccentricity planets and a long tail of moderately eccentric planets.}
	\label{fig:hab_eccs}
\end{figure}

Another interesting result we can take from our data is a distribution of all the maximum Earth eccentricities that are associated with systems with nonzero habitability probability. The histogram of these eccentricities is shown in Figure~\ref{fig:hab_eccs}.

\subsection{Discussion}\label{sec:griddisc}

\subsubsection{General Trends in One Dimension}\label{sec:griddiscgeneral}

At the level of flattening in Figure~\ref{fig:1D_violin}, much detail is lost, but some trends can be observed. First, it is clear that the most common outcome is no relative habitability. In fact, almost 60\% of the giant planet configurations have a relative habitability of zero. This is, in itself, not a noteworthy outcome, as the grid of parameter space was not chosen to necessarily be conducive to stable outcomes in every combination.

We note a trend of decreasing amounts of high relative habitability as the mass of the giant planets increase, with the greatest amount of relatively habitable systems occurring for the lowest $m_2$. For the locations of the giant planets, unsurprisingly when planet 1 is located in the HZ, relative habitability is extremely low. Also unsurprising is the reduction in relative habitability when the giant planets are close to one another at high $\alpha$.

The trend in eccentricity is a decreasing amount of high relative habitability system as eccentricity increases; however, this trend seems less strong than the trend in mass of the giant planets. Instead, it decreases gently and then falls off a ``cliff,'' past which there is almost no relative habitability. Indeed, for the case of $e_2~=~0.91$, there is not a single system with nonzero relative habitability. This is expected, as that high of an eccentricity causes orbit crossing even for the most widely separated giant planets.

\subsubsection{Coplanar Versus Inclined Trends in Two Dimensions}\label{sec:griddisccoplincl}

The effect of mild inclinations rather than coplanar systems is quite small, with the difference in mean relative habitability always less than 0.12 and typically much less. We do note a few features in Figure~\ref{fig:2D_incl}.

\begin{figure}
	\centering
    \includegraphics[width=\linewidth]{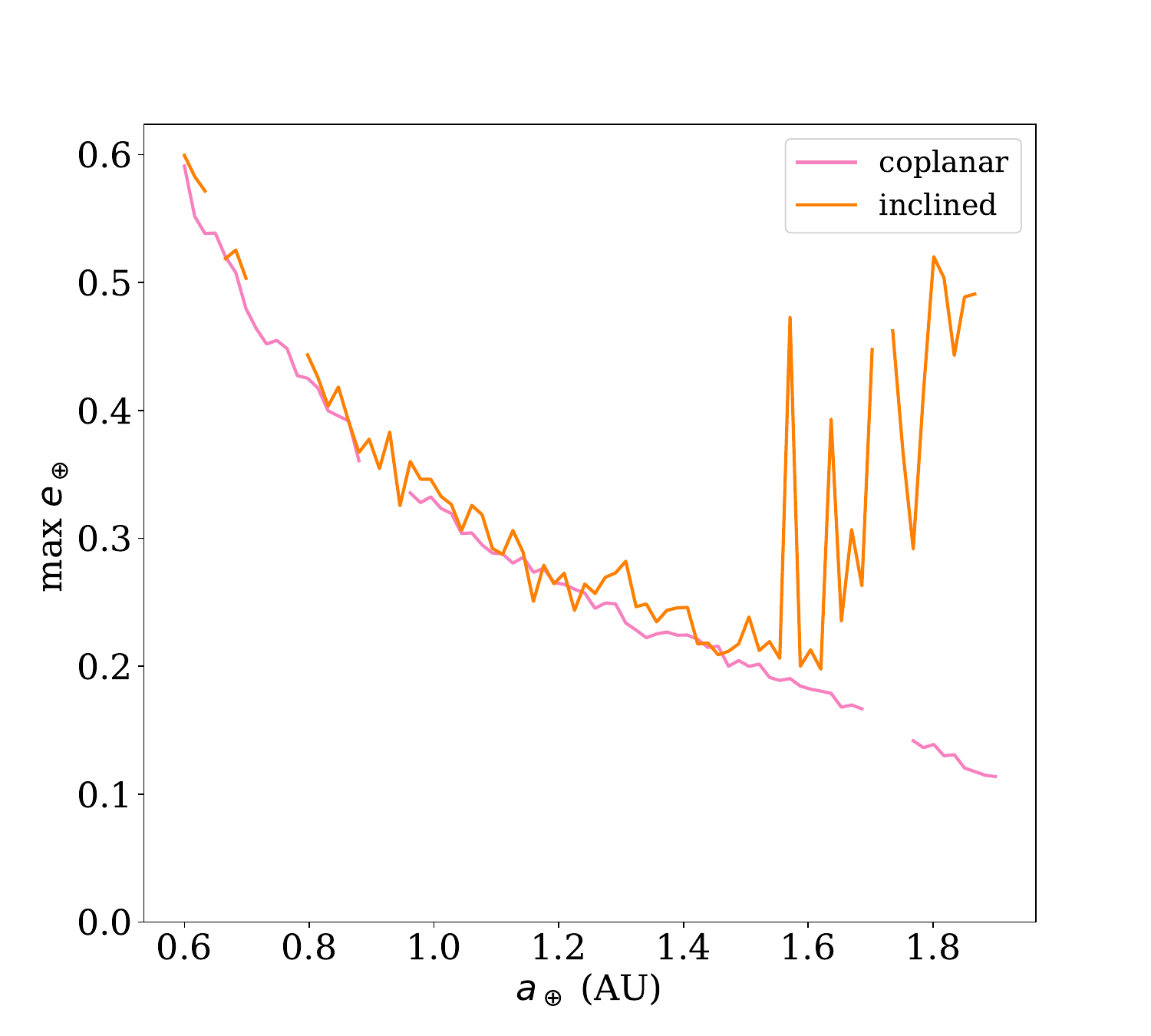}
	\caption[]{The maximum eccentricity of the Earths for two giant planet configurations. Systems with a stability outcome of zero are not plotted. Except for the inclination, the giant planet properties are the same in both configurations: $m_1$~=~$m_2$~=~10~$M_{Jup}$, $a_1$~=~10~AU, $\alpha$~=~0.354, $e_1$~=~$e_2$~=~0.370, $\Delta \varpi$~=~0. The coplanar configuration has a relative habitability of 0.716; the inclined configuration has a relative habitability of 0.844. The mildly excited eccentricities in the outer regions of the HZ creates a notable trend at high $a_1$ on Figure~\ref{fig:2D_incl}.}
	\label{fig:maxeE_inclcopl_higha1}
\end{figure}

In general, there is a slight preference for coplanar systems. However, this reverses for high $a_1$, where the inclined systems have higher mean relative habitability. This trend is strongest at middling masses, higher eccentricities, or middling $\alpha$. The increase in relative habitability is because a small amount of inclination can excite mild eccentricities in the Earths, and increasing the eccentricities of the Earth can extend the outer edge of the HZ and increase the relative habitability. When the giant planets are further out, this effect falls in the range of Earth semi-major axes that have the potential for a large increase in relative habitability as eccentricity increases. Also, because the planets are more widely separated, the systems are more likely to be stable even with excited eccentricities. An example of this phenomenon is illustrated in Figure~\ref{fig:maxeE_inclcopl_higha1}.

\begin{figure}
	\centering
    \includegraphics[width=\linewidth]{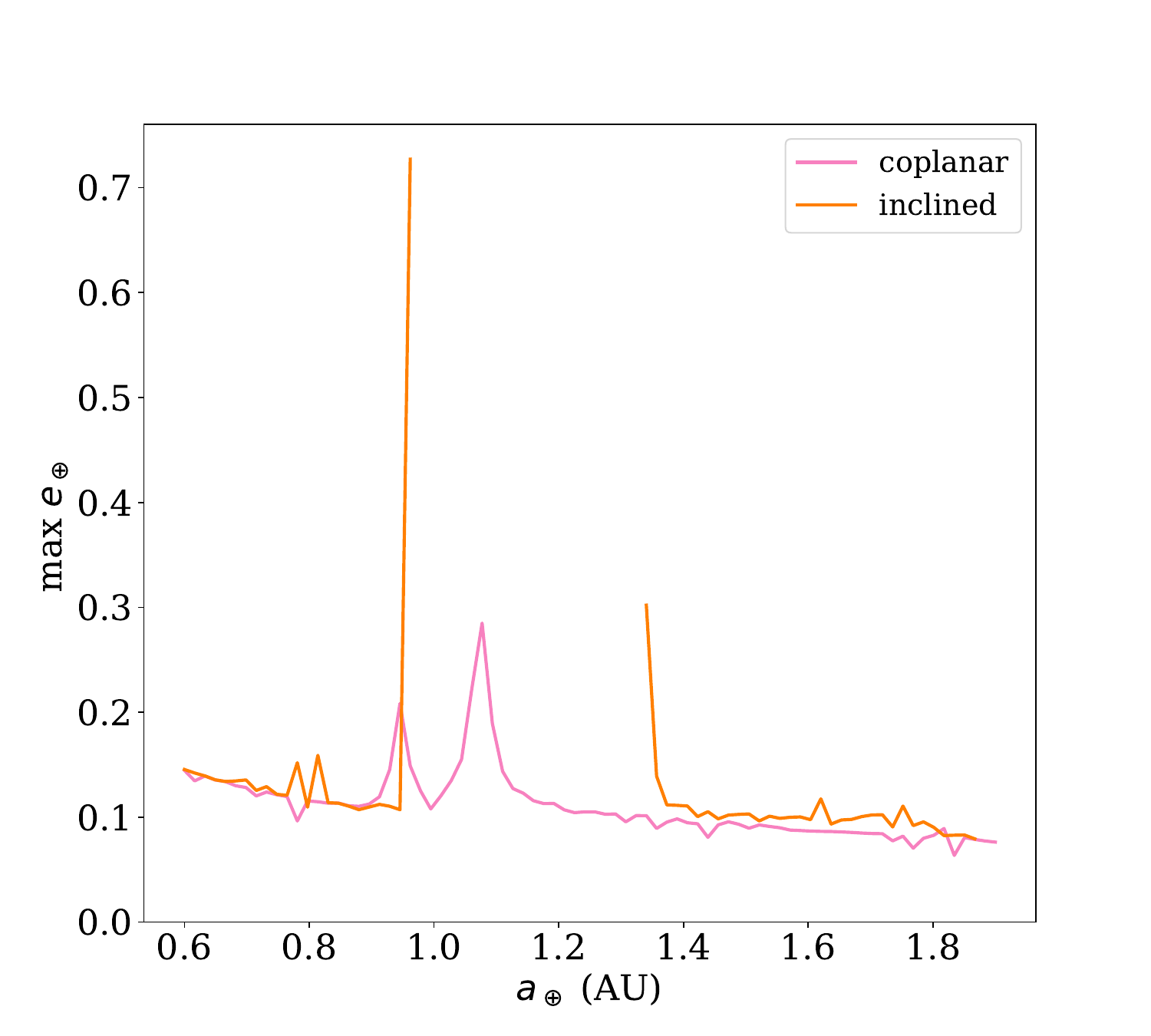}
	\caption[]{The maximum eccentricity of the Earths for two giant planet configurations. Systems with a stability outcome of zero are not plotted. Except for the inclination, the giant planet properties are the same in both configurations: $m_1$~=~10~$M_{Jup}$, $m_2$~=~3.4~$M_{Jup}$, $a_1$~=~4.37~AU, $\alpha$~=~0.428, $e_1$~=~0.025, $e_2$~=~0.061, $\Delta \varpi$~=~$\pi$. The coplanar configuration has a relative habitability of 0.837; the inclined configuration has a relative habitability of 0.418. The destabilizing effect of the inclination creates a notable feature at this $\alpha$/$a_1$ combination on Figure~\ref{fig:2D_incl}.}
	\label{fig:maxeE_inclcopl_alphaa1}
\end{figure}

Additionally, we note a strong preference (0.098 difference) for coplanar systems at $\alpha~=~0.43$ and $a_1~=~4.4$~AU. This difference is due to a large section of the center of the HZ that becomes unstable due to eccentricity excitation in the inclined systems but not in the coplanar systems. See Figure~\ref{fig:maxeE_inclcopl_alphaa1} for an illustration of this phenomenon. We note that the location of this excitation does not correspond with predicted secular resonances from Laplace-Lagrange theory (which are at 0.35 AU and 0.83 AU for eccentricity-pericenter and 0.95 AU for inclination-node) nor with any major mean-motion resonances lower than 5th order (6:1 between planet 1 and the Earth falls at 1.32 AU). It is possible that a higher-order secular resonance is at play; for example, the $[-1,1]$ resonance where a test particle has $-f+g_1$ frequency would fall at 1.06 AU. However, the exact details of this feature are beyond the scope of this work.

\subsubsection{Aligned Versus Antialigned Trends in Two Dimensions}\label{sec:griddiscalignment}

Similarly, the effect of the pericenter alignment is also quite small, though we note some features in Figure~\ref{fig:2D_align}. There is a clear trend with $\alpha$, where small $\alpha$ (more separated) shows a preference for antialigned systems while high $\alpha$ (more compact) shows a preference for aligned systems. This trend is clearer for higher masses of the giant planets. The likely explanation is simply that alignment of the pericenter can allow more tightly nested orbits.

\begin{figure}
	\centering
    \includegraphics[width=\linewidth]{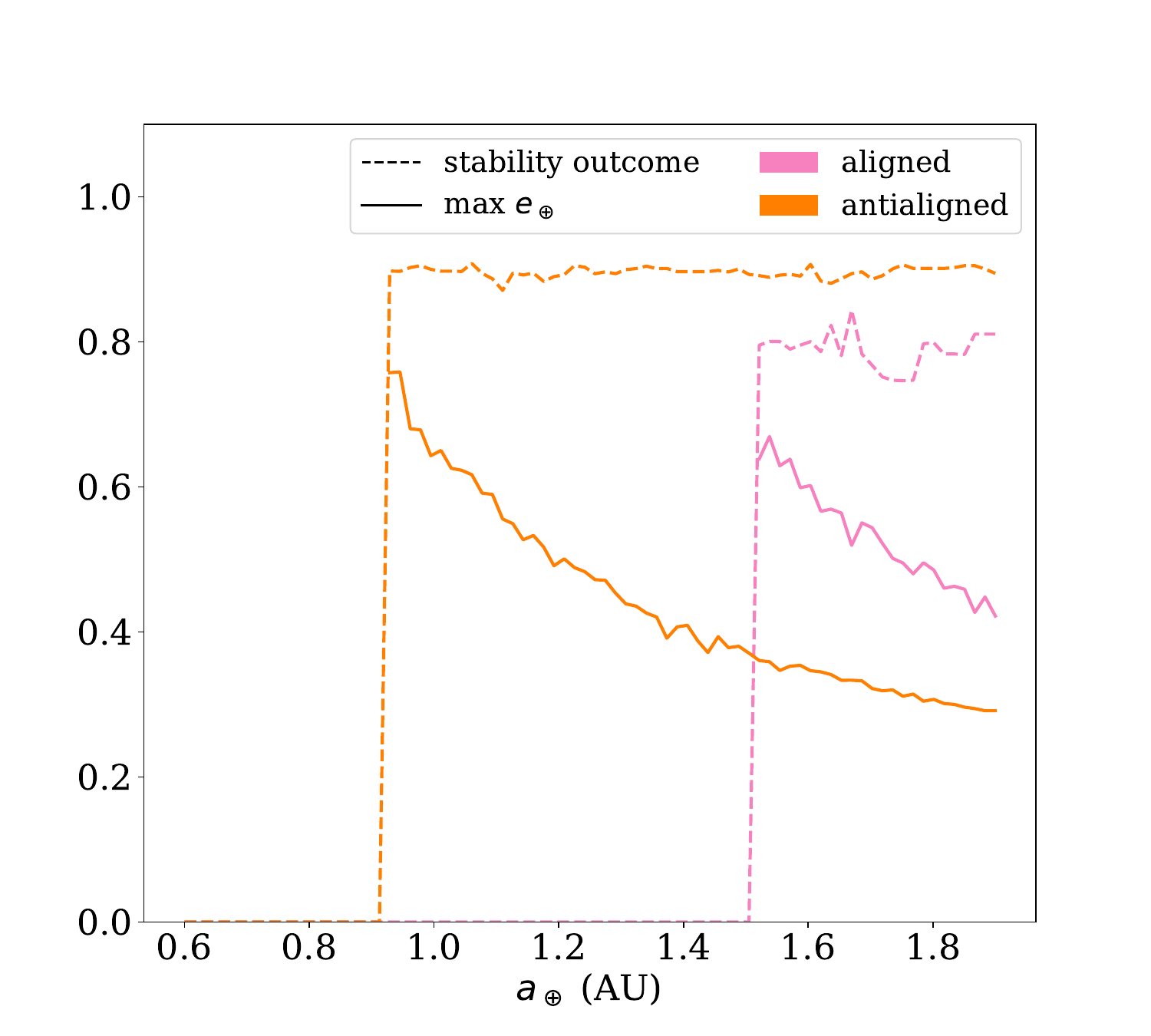}
	\caption[]{The stability outcome and maximum eccentricity of the Earths for two giant planet configurations. Eccentricities for systems with a stability outcome of zero are not plotted. Except for the alignment, the giant planet properties are the same in both configurations: $m_1$~=~$m_2$~=~10~$M_{Jup}$, $a_1$~=~10~AU, $\alpha$~=~0.241, $e_1$~=~$e_2$~=~0.025, $\Delta i$~=~0. The aligned configuration has a relative habitability of 0.297; the antialigned configuration has a relative habitability of 0.752. This is opposite the trend for similar configurations with lower giant planet masses, giving rise to the shift in the $m_1$/$m_2$ panel in Figure~\ref{fig:2D_align}.}
	\label{fig:outcome_maxeE_align_highm}
\end{figure}

There is also a small shift towards antialigned systems when the combined mass of the giant planets gets high. An example of this is illustrated in Figure~\ref{fig:outcome_maxeE_align_highm}, which shows the stability outcome and maximum eccentricity of the Earths for aligned and antialigned versions of a giant planet configuration. The instability zone for the aligned configuration extends further into the HZ. For similar systems with less combined mass in the giant planets, the opposite trend holds. Interestingly, the giant planets in the example in Figure~\ref{fig:outcome_maxeE_align_highm} could be considered observable by direct imaging techniques \citep{2010Traub}, indicating that we can potentially make habitability inferences for such systems.

We note two strong areas of preference for coplanar systems when $e_1$ or $e_2$ is 0.15 and $\alpha$~=~0.52. There seem to be several factors contributing to this trend. For example, in one configuration with $m_1$~=~6.7~$M_{Jup}$, $m_2$~=~0.1~$M_{Jup}$, $a_1$~=~0.1~AU, $\alpha$~=~0.518, $e_1$~=~0.061, $e_2$~=~0.150, and $\Delta i$~=~0.222, there are no unstable systems, but the aligned case simply has consistently higher stability outcomes predicted by \texttt{SPOCK}, leading to a relative habitability of 0.917 for the aligned system and 0.558 for the antialigned system. In another configuration with $m_1$~=~$m_2$~=~0.1~$M_{Jup}$, $a_1$~=~4.365~AU, $\alpha$~=~0.518, $e_1$~=~0.150, $e_2$~=~0.025, and $\Delta i$~=~0, there appear to be eccentricity resonances that excite the eccentricity of the Earth in approximately equal magnitude in both cases, but the one at approximately 1.8~AU results in instability for the antialigned case but not the aligned case, leading to a relative habitability of 1.023 for the aligned configuration and 0.839 for the antialigned configuration.

\subsubsection{Eccentricity Distribution of Potentially Habitable Earths}\label{sec:griddisceccs}

Considering the maximum eccentricity distribution of the possibly-habitable Earths (Figure~\ref{fig:hab_eccs}), we note two components of the distribution. The first is a large excess of low-eccentricity planets and the second is a long tail of moderately excited eccentricities. The low-eccentricity planets likely benefit from increased stability and a location well within the HZ, while some of the moderately eccentric planets are likely located exterior to the circular HZ and receive a boost in potential habitability thanks to their excited eccentricities. However, too much eccentricity excitation can also lead to instability, so very high eccentricities are rarely encountered.

\subsubsection{Parameter Sensitivity}\label{sec:griddiscsensitivity}

One question we might hope to answer with this data: which of the varied parameters has the largest effect on the relative habitability of a system? The problem of analyzing a multidimensional function is a complex one, and here we apply a rather simple analysis to help us better understand the interplay between parameters.

\begin{figure}
	\centering
    \includegraphics[width=\linewidth]{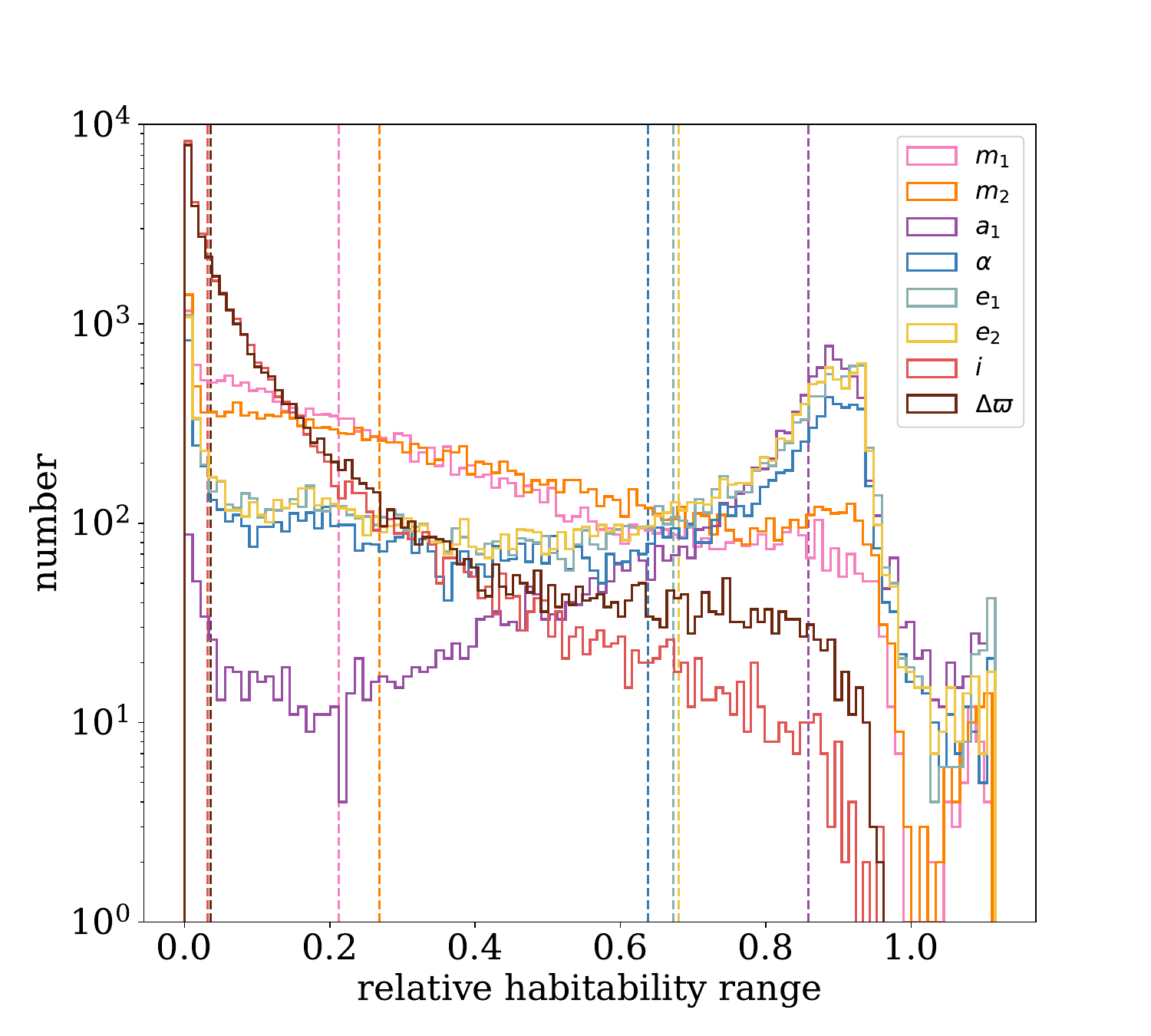}
	\caption[]{The distribution of nonzero relative habitability ranges for each parameter, calculated for every combination of the other parameters. The median values are indicated by a dashed vertical line.}
	\label{fig:ptp}
\end{figure}

For each parameter, we consider the function of relative habitability while each other parameter is held constant. From that one-dimensional function, we calculate the range of relative habitabilities (the maximum minus the minimum). This calculation is repeated for each possible combination of the other parameters. The distribution and median of these relative habitability ranges are shown in Figure~\ref{fig:ptp} for each of the parameters. For visibility, we've considered only the nonzero ranges, but this does not qualitatively change the conclusion.

A higher range is associated with more change from that parameter, so we identify the parameters from most to least impactful based on their median values: $a_1$, $e_2$, $e_1$, $\alpha$, $m_2$, $m_1$, $\Delta \varpi$, $i$.

\subsubsection{The Most Habitable Configurations}

\begin{figure*}
	\centering
    \includegraphics[width=\linewidth]{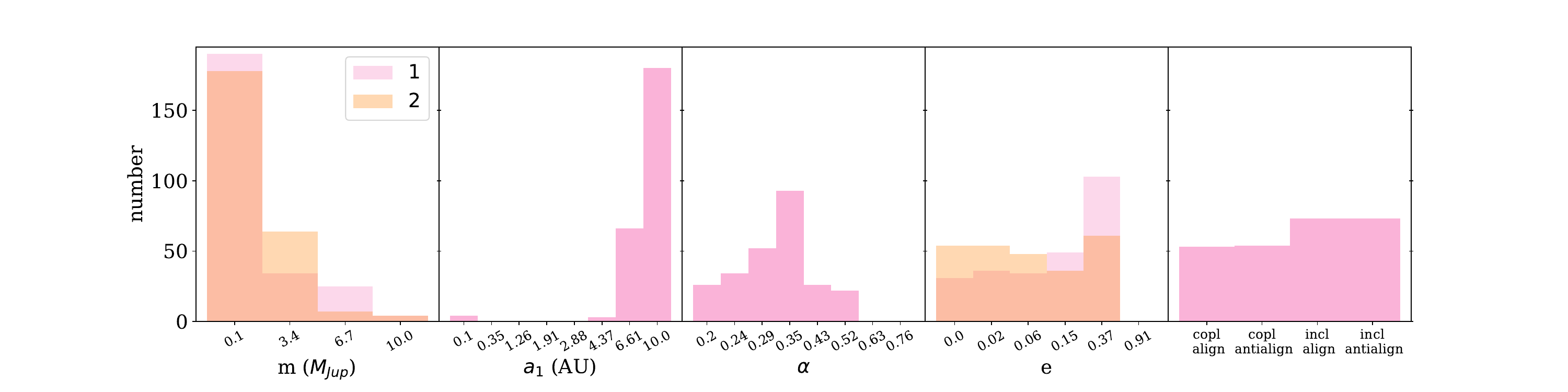}
	\caption[]{The distribution of systems with a relative habitability greater than one. The initial grid is uniform in each parameter; here, clear features are seen.}
	\label{fig:highest_relhab}
\end{figure*}

There are a few systems that have relative habitabilities greater than one. This means that having giant planets in that configuration makes it \textit{more} likely that an Earth analog could be habitable than a system with only the Earth analog. There are 253 such configurations. In Figure~\ref{fig:highest_relhab}, the histograms for each parameter are shown. Each parameter started with a completely uniform distribution, but we see that there is a strong selection effect when looking at these ultra-habitable systems.

The mechanism that makes it possible for a configuration to have a relative habitability greater than one is eccentricity excitation of the Earth. Based on our habitability model, Earths that lie external to the circular HZ can become potentially habitable with excited eccentricity. If this effect is not offset by instability or the loss of the inner portion of the HZ (where excited eccentricity reduces the potential habitability), the configuration can have a relative habitability greater than one (i.e., considered more habitable than a system with an Earth as the only planet). Most of these effects were seen in our previous results, though they become more distinct here. 

First, it is clear that these systems tend to have lower masses in both giant planets. More massive planets likely induce additional instabilities that offset the gains in relative habitability from eccentricity excitation. Similarly, very compact (high $\alpha$) configurations are not seen, and there is a clear preference for the middling  $\alpha$~=~0.35.

For the location of the giant planets, these ultra-habitable systems tend to be located further away from the HZ. This location preference is likely because more distant planets are less likely to destabilize the Earths while also being located to excite Earth eccentricities for planets located exterior to the HZ, where there is the most potential for increasing the habitability probability. This location preference also likely gives rise to the peak in the $\alpha$ distribution, as when $a_1$~=~10~AU, Laplace-Lagrange theory puts the secular resonance locations near the outer edge of the HZ for $\alpha$~=~0.35 ($a_2$~=~28.3~AU). When $a_1$~=~6.61~AU, that shifts towards $\alpha$~=~0.43 ($a_2$~=~15.4~AU), which also is represented in Figure~\ref{fig:highest_relhab}.

For eccentricity, there are no highly eccentric systems, which is unsurprising as very few of these systems have any stable configurations at all. There is a marked difference in the eccentricity distributions between the two giant planets. For planet 2, the distribution is relatively uniform up to the stability cliff. For planet 1, there is a clear preference for moderate eccentricities. Again, this is likely the most efficient way to transfer eccentricity to the Earths (which all begin on circular orbits) without destabilizing the systems.

Lastly, there is almost no difference between the aligned or anti-aligned pericenters. There is a small but clear preference for inclined systems over coplanar ones; this preference makes sense given that even a small amount of inclination can lead to eccentricity excitation.

Although the details of these ultra-habitable parameters are influenced by our choice of habitability model, it seems likely that exterior giant companions can increase the size of the HZ around Sun-like stars, particularly when these giant companions are well-separated cold sub-Jupiters with moderate eccentricities and inclinations.

\subsubsection{Giants in the Habitable Zone}

One would expect that positioning a pair of giant planets right in the HZ would eliminate the possibility for habitable Earths. And while we do see a significant reduction of relative habitability for those configurations, it is not a complete gap. Our method precludes the possibility of Trojan Earths orbiting in a 1:1 resonance with the giant planets, as any orbit crossing is considered unstable, so though these types of planets may exist \citep{2004Dvorak}, they cannot explain the result here.

\begin{figure*}
	\centering
    \includegraphics[width=\linewidth]{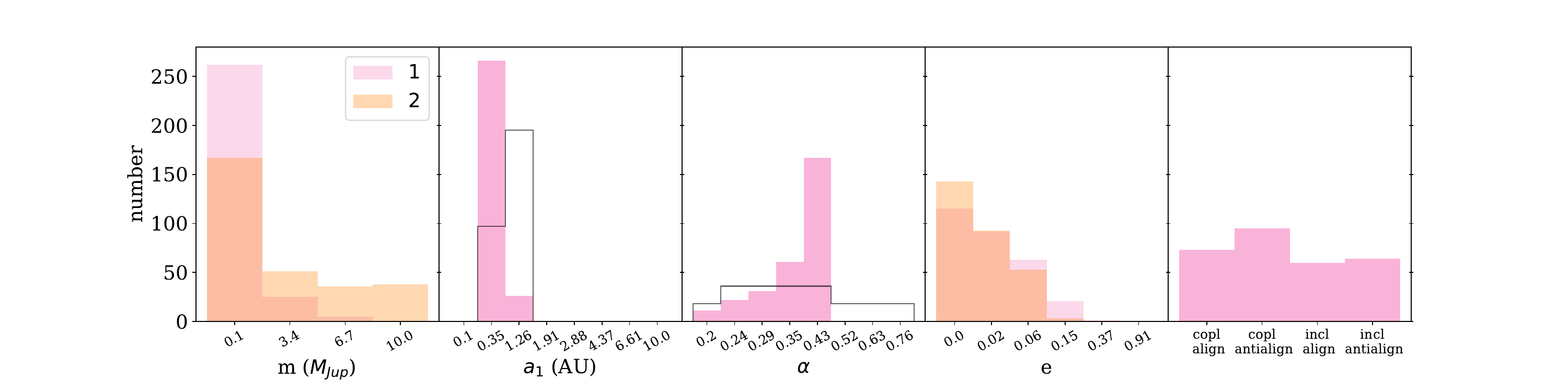}
	\caption[]{The distribution of systems with a relative habitability greater than 0.5 that overlap the circular HZ. The initial distribution of overlapping configurations (without applying the relative habitability criterion) is uniform in each parameter except for $a_1$/$\alpha$, where they are shown in black for comparison. Clear features arise when considering the configurations with reasonable relative habitability.}
	\label{fig:overlap_HZ}
\end{figure*}

Let us consider the configurations where either giant planet falls between the most optimistic boundaries of the HZ, that is, between 0.75~AU and 1.77~AU. This is when $a_1$~=~1.26~AU, with any $\alpha$, or when $a_1$~=~0.35~AU and $\alpha$~=~0.24, 0.29, 0.35, or 0.43. Then we look at the configurations that have a relative habitability greater than 0.5, giving us a total of 292 configurations.  In Figure~\ref{fig:overlap_HZ}, the histograms for each parameter are shown. Each parameter started with a completely uniform distribution (except for $a_1$/$\alpha$ as mentioned; their distributions are shown in black in Figure~\ref{fig:overlap_HZ} for comparison), but we see clear trends when applying the relative habitability criterion.

We first note a strong preference for low $m_1$ but only a middling preference for low $m_2$. It is also far more likely that the relative habitability stays high when it is planet 2 that overlaps the HZ rather than planet 1, despite the fact that these systems are outnumbered two to one in the sample of overlapping configurations. This is the case even for the overlapping $a_2$ closest to the overlapping $a_1$ (when $\alpha$~=~0.29, $a_2$~=~1.21~AU). For example, the highest relative habitability is 0.86 when planet 2 is the overlapping planet (0.66 for $a_2$~=~1.21~AU) but only 0.55 when planet 1 is the overlapping planet. It is possible to have both planets overlapping the HZ when $a_1$~=~1.26~AU and $\alpha$~=~0.76, but none of these systems have a relative habitability higher than 0.13.

There is a peak in the $\alpha$ distribution at $\alpha$~=~0.43. When $a_1$~=~0.35~AU, that puts $a_2$ at 0.814~AU, just barely overlapping the inner edge of the HZ. As $\alpha$ decreases, planet 2 moves further into the HZ and decreases the number of configurations with high relative habitability.

Unsurprisingly, we see a strong preference for lower eccentricities, slightly more marked for $e_2$. Given that planet 2 is closer to the HZ and the compact nature of these systems, large amounts of eccentricity can quickly cause instability.

Lastly, we see a slight preference for the coplanar and antialigned configurations. Coplanarity preference is expected, as even a small amount of inclination can provide some additional instability, but the preference for anti-alignment is somewhat unexpected, as these compact systems would see stability benefits from an aligned orientation. However, given that most systems have very small eccentricities, the pericenter orientation becomes less important, so it is possible that this feature is insignificant. Assuming $\sqrt(N)$ error, the difference between aligned (133) and anti-aligned (159) is $\sim1.5\sigma$ discrepant.

From these results, we can conclude that the HZ is wide enough to accommodate both a habitable Earth and a giant planet, though likely not two giant planets. These systems are more possible with smaller giant planets, particularly the inner giant planet, and with low eccentricities.

We do note here that this discussion does not include the likelihood of forming such a system in the first place. Giant planets have a strong influence on the formation of terrestrial planets \citep{2012Morbidelli,2019Childs}. Effects from the in situ formation of giants in or near the HZ or from their migration to that location could limit the possibility of forming or migrating Earths into the HZ \citep{2006Raymond,2017Darriba}; inclusion of these effects are left for another study.

\subsection{Future Work}\label{sec:future}

This paper puts forth a process for evaluating system architectures for habitability and creates a rich initial dataset for examining the impact of giant companions on the habitability of an Earth. There are many important steps that could be taken to iterate and improve upon what is presented here.

We've seen that dynamical interactions with giant planets can affect the eccentricity of an Earth analog and therefore its potential habitability. Similarly, the system architecture can affect the obliquity of Earth-like planets, which also strongly influences the climate \citep{2015Linsenmeier, 2017Kane}.

We use a very simplified and discontinuous habitability model. More accurate habitability zone limits taking into account additional 1-D and 3-D climate model results could improve this model; multiple models could even be used to compare and contrast. The habitability model could even incorporate more statistical methods to better match our current knowledge of the HZ \citep{2015Zsom,2021Mendez}.

We consider only exactly Earth-like planets, ignoring possible effects of terrestrial planet mass, rotation rate, magnetic field, density, and composition, all of which have enormous implications for climate and habitability \citep{2014Yang,2018Meadows}.

We consider only Sun-like stars. The HZ varies significantly around different stellar types, due to differences in temperature and spectrum \citep{2013Kopparapu}, and many dynamical effects scale with the planet-star mass ratio. Additionally, a large portion of stars are found in binaries, which can effect habitability for both circumstellar and circumbinary planets \citep{2014Cuntz,2021Georgakarakos}. Stars also change in luminosity over time, which can affect the location of the HZ \citep{2015Baraffe}.

We achieve only mild resolution in each parameter considered, likely missing narrow features in our multidimensional grid, as we see such features arise in our fiducial system. More computational time is required to achieve better resolution.

We neglect the dynamical histories of the systems and constraints from formation. It is likely that many systems with giant planets undergo instabilities \citep{2008Juric}, which could affect both the formation of terrestrial planets and the stability of those planets, depending on the timescales of instability and formation. Furthermore, we consider any instability within the system to preclude habitability, which may not necessarily be the case \citep{2020Kokaia}. Some of our giant planet configurations may not be compatible with the formation of terrestrial planets or may be difficult to explain in and of themselves (e.g. systems with $e_1$~=~0.91 and $e_2$~=~0). We also do not consider the possibility of overlapping but stable orbits, such as Pluto and Neptune have in the solar system, nor of co-orbital planets. Neither do we include tidal effects or circularization, which could be significant for some of our planets \citep{1998Eggleton}.

We also do not consider multiple terrestrial planets. Multiple terrestrial planets might lead to additional instability due to interactions with one another, or they may in fact stabilize one another, such as the Venus-Earth interaction in our solar system that displaces Venus's eccentricity from a secular resonance \citep{1998Innanen}.

\section{Conclusions}\label{sec:conclusion}

Determining the habitability of an exoplanet requires taking into account the interplay of a large number of complex and interdisciplinary factors, many of which are not yet well understood. In this paper, we investigated one aspect of habitability external to the planet itself: the configuration of its giant companions. The system architecture relates to habitability via the destabilization of planets in the habitable zone as well as the excitation of those planets' eccentricities.

From our one-dimensional analysis of two fiducial systems (one exterior and one interior), we saw that the giant planets can have very large and sharp effects as they get close to one another or to the HZ and when their eccentricities are high. When the giants are exterior to the HZ, Kozai-Lidov cycles can destabilize Earths in the HZ for highly mutually inclined systems. Resonances can provide islands of protection at high eccentricities. More mild effects arise as the location of secular resonances shift throughout the system, such as a dependence on the mass ratio of the giant planets.

We undertook a truly multidimensional investigation of system architectures and determined that the presence of giant companions usually decreases the relative habitability of a system compared to the presence of an Earth-like planet alone. This is particularly true when the giant planets are very massive, close to the habitable zone, tightly spaced, and highly eccentric, though exceptions exist, such as the coexistence of giant and Earth-like planets within the HZ. However, the giant planets can sometimes \textit{increase} the relative habitability of a system by providing mild eccentricity excitation that extends the outer edge of the HZ, usually when the giant companions are on wide, mildly eccentric, and inclined orbits.

From our results, we see that the location of the inner giant planet has the most effect on the relative habitability of a system, followed by several parameters of similar strength with the outer giant planet's eccentricity, the inner giant planet's eccentricity, and the semi-major axis ratio of the giant planets, while the mass of the outer giant planet and then the mass of the inner giant planet have the least effect of the main six parameters, aside from alignment and inclination which are not well-sampled.

There are many avenues to build upon the work presented here. Even so, we have created a rich dataset, and our results illuminate several of the many possible effects of the system architecture on the habitability of exoplanets.

\section*{Acknowledgements}

Thanks to Dorian Abott and Xuan Ji for many helpful discussions regarding eccentricity and habitability. This work was completed in part with resources provided by the University of Chicago Research Computing Center. Simulations in this paper made use of the \texttt{REBOUND} code which can be downloaded freely at http://github.com/hannorein/rebound. DF acknowledges support of grant NASA-NNX17AB93G through NASA's Exoplanet Research Program.

%%%%%%%%%%%%%%%%%%%%%%%%%%%%%%%%%%%%%%%%%%%%%%%%%%
\section*{Data Availability}

The data used in this paper is available for download at http://dx.doi.org/10.5281/zenodo.6324216.



\bibliographystyle{mnras}
\bibliography{archhab} % if your bibtex file is called example.bib

%\begin{thebibliography}{99}
%\end{thebibliography}

%%%%%%%%%%%%%%%%%%%%%%%%%%%%%%%%%%%%%%%%%%%%%%%%%%

%%%%%%%%%%%%%%%%% APPENDICES %%%%%%%%%%%%%%%%%%%%%

\appendix

\section{Stability Outcome Results}\label{app:stability_results}

We include here Table~\ref{apptab:grid_stability_results} with the complete results for each three-planet system (each giant planet configuration and its associated 80 Earth locations). We applied our habitability model (Section~\ref{sec:hab}) to these results to determine our relative habitability results (Table~\ref{tab:grid_results}); we include these intermediate results to allow for the potential application of different habitable models or for context in understanding the results in Table~\ref{tab:grid_results}. The ``outcome codes'' in Table~\ref{apptab:grid_stability_results} refer to which step of the process outlined in Section~\ref{sec:process} resulted in an outcome of zero:

\begin{itemize}
  \item 2: 2 giant planets fail analytical stability criterion
  \item o: orbit crossing is predicted by Laplace-Lagrange theory
  \item z: zero stability predicted by \texttt{SPOCK}
  \item u: unstable during the $5\times10^6$-orbit integration
  \item h: high spectral fraction predicts instability
\end{itemize}

\begin{table*}
    \centering
    \caption[]{Stability results for a multidimensional set of giant planet parameters and Earth locations. See the text for details on the outcome codes.}
    \label{apptab:grid_stability_results}
    \resizebox{\textwidth}{!}{%
    \begin{tabular}{cccccccccccc}
        \hline
        $m_1$ & $m_2$ & $a_1$ & $\alpha$ & $e_1$ & $e_2$ & $\Delta i$ & $\Delta \varpi$ & $a_\oplus$ & Outcome Code & Outcome & Max $e_\oplus$ \\
        ($M_{Jup}$) & ($M_{Jup}$) & (AU) &  &  &  & (rad) & (rad) & (AU) &  &  &  \\ \hline
        6.7 & 10 & 10.000000000000000 & 0.427985022944864 & 0.149968483550237 & 0.369828179780266 & 0.000000000000000 & 0.000000000000000 & 1.554430379746840 & z & 0.000000000000000 & 0.000000000000000 \\
        0.1 & 10 & 1.258925411794170 & 0.241466642166516 & 0.024660393372343 & 0.912010839355910 & 0.221656815003280 & 3.141592653589790 & 1.208860759493670 & 2 & 0.000000000000000 & 0.000000000000000 \\
        6.7 & 0.1 & 10.000000000000000 & 0.427985022944864 & 0.024660393372343 & 0.000000000000000 & 0.000000000000000 & 3.141592653589790 & 1.669620253164560 &  & 0.926216959953308 & 0.100342776605637 \\
        10 & 10 & 1.905460717963250 & 0.292222926488148 & 0.000000000000000 & 0.369828179780266 & 0.000000000000000 & 0.000000000000000 & 1.455696202531650 & o & 0.000000000000000 & 0.000000000000000 \\
        6.7 & 3.4 & 4.365158322401660 & 0.758577575029184 & 0.912010839355910 & 0.912010839355910 & 0.221656815003280 & 3.141592653589790 & 1.636708860759490 & z & 0.000000000000000 & 0.000000000000000 \\
        10 & 0.1 & 0.354813389233575 & 0.241466642166516 & 0.149968483550237 & 0.060813500127872 & 0.000000000000000 & 0.000000000000000 & 1.801265822784810 & z & 0.000000000000000 & 0.000000000000000 \\
        10 & 6.7 & 1.905460717963250 & 0.427985022944864 & 0.912010839355910 & 0.369828179780266 & 0.221656815003280 & 3.141592653589790 & 0.978481012658228 & 2 & 0.000000000000000 & 0.000000000000000 \\
        0.1 & 10 & 2.884031503126610 & 0.517947467923121 & 0.060813500127872 & 0.000000000000000 & 0.221656815003280 & 3.141592653589790 & 1.521518987341770 &  & 0.567690968513489 & 0.147312329046831 \\
        0.1 & 10 & 2.884031503126610 & 0.241466642166516 & 0.912010839355910 & 0.369828179780266 & 0.000000000000000 & 3.141592653589790 & 1.554430379746840 & 2 & 0.000000000000000 & 0.000000000000000 \\
        0.1 & 3.4 & 2.884031503126610 & 0.241466642166516 & 0.369828179780266 & 0.912010839355910 & 0.000000000000000 & 3.141592653589790 & 1.143037974683540 & o & 0.000000000000000 & 0.000000000000000 \\
        3.4 & 0.1 & 10.000000000000000 & 0.292222926488148 & 0.149968483550237 & 0.024660393372343 & 0.000000000000000 & 3.141592653589790 & 1.175949367088610 &  & 0.935216546058655 & 0.109784552928324 \\
        0.1 & 10 & 10.000000000000000 & 0.199526231496888 & 0.024660393372343 & 0.000000000000000 & 0.000000000000000 & 0.000000000000000 & 1.406329113924050 &  & 0.904384553432465 & 0.092353750797595 \\
        0.1 & 3.4 & 6.606934480075960 & 0.758577575029184 & 0.060813500127872 & 0.149968483550237 & 0.221656815003280 & 3.141592653589790 & 0.797468354430380 & 2 & 0.000000000000000 & 0.000000000000000 \\
        0.1 & 6.7 & 10.000000000000000 & 0.241466642166516 & 0.149968483550237 & 0.060813500127872 & 0.221656815003280 & 3.141592653589790 & 1.686075949367090 &  & 0.878525078296661 & 0.124453378884910 \\
        10 & 10 & 4.365158322401660 & 0.199526231496888 & 0.369828179780266 & 0.024660393372343 & 0.221656815003280 & 3.141592653589790 & 1.784810126582280 & z & 0.000000000000000 & 0.000000000000000 \\
        0.1 & 3.4 & 1.258925411794170 & 0.241466642166516 & 0.000000000000000 & 0.912010839355910 & 0.000000000000000 & 3.141592653589790 & 1.406329113924050 & z & 0.000000000000000 & 0.000000000000000 \\
        6.7 & 6.7 & 10.000000000000000 & 0.353648180962444 & 0.000000000000000 & 0.369828179780266 & 0.221656815003280 & 3.141592653589790 & 1.472151898734180 & u & 0.000000000000000 & 0.000000000000000 \\
        10 & 10 & 0.354813389233575 & 0.517947467923121 & 0.024660393372343 & 0.000000000000000 & 0.000000000000000 & 3.141592653589790 & 1.241772151898730 & u & 0.000000000000000 & 0.000000000000000 \\
        6.7 & 6.7 & 4.365158322401660 & 0.626820017397041 & 0.912010839355910 & 0.024660393372343 & 0.000000000000000 & 0.000000000000000 & 0.781012658227848 & 2 & 0.000000000000000 & 0.000000000000000 \\
        3.4 & 10 & 1.905460717963250 & 0.427985022944864 & 0.149968483550237 & 0.912010839355910 & 0.000000000000000 & 0.000000000000000 & 1.274683544303800 & o & 0.000000000000000 & 0.000000000000000 \\
        \hline
        \multicolumn{12}{c}{Note:  The full version of Table~\ref{apptab:grid_stability_results} is available for download at http://dx.doi.org/10.5281/zenodo.6324216 \citep{datasets}.} \\
        \multicolumn{12}{c}{A random subset of rows are shown here for guidance regarding its form and content.}
    \end{tabular}%
    }
\end{table*}

%%%%%%%%%%%%%%%%%%%%%%%%%%%%%%%%%%%%%%%%%%%%%%%%%%

% Don't change these lines
\bsp	% typesetting comment
\label{lastpage}
\end{document}